%% file: main.tex
\documentclass{article}
\usepackage[margin=0.5in]{geometry}
\usepackage[american]{babel}
\usepackage{csquotes}
\usepackage[style=vancouver,maxcitenames=1,uniquelist=false,citestyle=numeric-comp]{biblatex}
\renewcommand{\thetable}{\Roman{table}}
\usepackage{float}
\usepackage{xcolor}
\usepackage{listings}
\usepackage[normalem]{ulem}
\useunder{\uline}{\ul}{}

\usepackage{tabularx}
\usepackage{booktabs,graphicx, wrapfig, subcaption,caption,tabularx}
\usepackage{adjustbox}
\renewcommand{\thetable}{\arabic{table}}
\usepackage[T1]{fontenc}
\usepackage[font=small, labelfont=bf]{caption}

\begin{document}

\title{Generative artificial intelligence in ophthalmology: multimodal retinal images for the diagnosis of Alzheimer’s disease with convolutional neural networks}
%
%
%
\author{Slootweg, I.R.
Thach, M.
Curro-Tafili, K.R. 
Verbraak, F.D.
Bouwman, F.H. 
Pijnenburg, Y.A.L.  \\
Boer, J.F. de  
Kwisthout, J.H.P.
Bagheriye, L. 
González, P.J.  
}

\maketitle

\section{Synopsis}
\input{Synopsis}
\section{Abstract}
\input{Abstract}
\section{Introduction}
\input{Introduction/Introduction}
\section{Materials and Methods}
\input{Methods/Datasets}

\input{Methods/Models}


\section{Results}
\input{Results/Synthesis}
\input{Results/Classification}
\input{Results/Cams}

\section{Discussion}
\input{Discussion/Discussion}

\newpage
\clearpage
\printbibliography
\newpage
\section*{Appendix}
\begin{refsection}
\setcounter{table}{0}
\renewcommand{\thetable}{S\arabic{table}}
\setcounter{figure}{0}
\renewcommand{\thefigure}{S\arabic{figure}}
\input{Supplementary/Dataset}

\input{Supplementary/Models}
\input{Supplementary/Cams}
\newpage
\printbibliography[heading=subbibliography]

\end{refsection}
\newpage

\end{document}

%% file: Synopsis.tex

Denoising diffusion probabilistic models can generate retinal images suitable for augmenting training datasets. Deep learning shows promise for non-invasive Amyloid Positron Emission Tomography status prediction, with best performance using multimodal retinal imaging and metadata.

%% file: Abstract.tex
\textbf{Background/Aim:}
This study aims to predict Amyloid Positron Emission Tomography (AmyloidPET) status with multimodal retinal imaging and convolutional neural networks (CNNs) and to improve the performance through pretraining with synthetic data. \\
\textbf{Methods} Fundus autofluorescence, optical coherence tomography (OCT), and OCT angiography images from 328 eyes of 59 AmyloidPET positive subjects and 108 AmyloidPET negative subjects were used for classification. Denoising Diffusion Probabilistic Models (DDPMs) were trained to generate synthetic images and unimodal CNNs were pretrained on synthetic data and finetuned on real data or trained solely on real data. Multimodal classifiers were developed to combine predictions of the four unimodal CNNs with patient metadata. Class activation maps of the unimodal classifiers provided insight into the network's attention to inputs. \\
\textbf{Results} DDPMs generated diverse, realistic images without memorization. Pretraining unimodal CNNs with synthetic data improved AUPR at most from $0.350$ to $0.579$. Integration of metadata in multimodal CNNs improved AUPR from $0.486$ to $0.634$, which was the best overall best classifier. Class activation maps highlighted relevant retinal regions which correlated with AD.\\
\textbf{Conclusion} Our method for generating and leveraging synthetic data has the potential to improve AmyloidPET prediction from multimodal retinal imaging. A DDPM can generate realistic and unique multimodal synthetic retinal images. Our best-performing unimodal and multimodal classifiers were not pretrained on synthetic data, however pretraining with synthetic data slightly improved classification performance for two out of the four modalities. \\ \\
\textbf{What is already known on this topic}Alzheimer’s disease (AD) is diagnosed through invasive and expensive methods
such as Amyloid Positron Emission Tomography (AmyloidPET) and cerebrospinal fluid analysis. 
The retina shows potential for non-invasive AD diagnostics with convolutional neural networks through imaging techniques like optical coherence tomography (OCT), OCT  angiography and Fundus Scanning Laser Opthalmoscopy (CNNs) but this is limited by the availability of few and small datasets. \\
\textbf{What this study adds}
This study demonstrates that synthetic multimodal retinal images generated by a DDPM are suitable for augmenting training datasets and pretraining CNNs on synthetic data can enhance the diagnostic accuracy for AmyloidPET.\\
\textbf{How this study might affect research, practice, or policy}
The findings suggest that generative AI holds promise for improving non-invasive AmyloidPET diagnosis. In turn, this allos for community-based AD screening, and offers av cost-effective and accessible alternative to current methods.


%% file: Introduction/Introduction.tex
Alzheimer's disease (AD) is a progressive neurodegenerative condition that starts years before symptoms appear and poses an important public healthcare concern due to aging populations worldwide.\parencite{Rasmussen2019} The disease is currently irreversible. Therefore, AD drug research is focused on early treatment to delay or prevent the development into dementia.\parencite{Wu2023-je} 
Current clinical diagnosis relies on detection of decreased amyloid-beta levels, increased total tau and phosphorylated tau levels in the cerebrospinal fluid as well as imaging amyloid protein deposition in the brain using an Amyloid Positron Emission Tomography (AmyloidPET).\parencite{McKhann2011} These techniques, however, are expensive and invasive and not practical for community-based screening for early onset of AD.\newline \newline 
The retina is derived from the same embryological tissue as the brain and easily accessible for non-invasive screening via widely available imaging techniques such as Optical Coherence Tomography (OCT), OCT Angiography (OCT-A) and Fundus Scanning Laser Ophthalmoscopy photography (FSLO). Studies have associated retinal imaging parameters with AD.\parencite{Zhang2022, Cheung2021,Zabel2021,Jin2021,Salazar2022} These studies revealed abnormalities in the retinal nerve fiber layer, blood vessels and the optic nerve that correlate with changes in the brain of patients with AD.
\newline \newline 
Advances in artificial intelligence (AI) have allowed for the development of parameter-based and image-based classification models for the prediction of AD. Most studies focus on brain MRI and rarely on retinal imaging-based classification.\parencite{Basaia2019, Amoroso2018, Wang2022} In turn, inputs of most retinal imaging-based implementations rely on the selection and manual extraction of image-derived features which can introduce information loss. However, several studies tried to mitigate these problems by developing convolutional neural networks (CNNs) for AD-related predictions with multimodal retinal images. But adding retinal imaging (AUROC = $0.836$) to an existing set of inputs encompassing ganglion cell-inner plexiform layer maps, quantitative data and patient data (AUROC = $0.841$) from $284$ eyes of $159$ subjects, did not improve performance of AD detection.\parencite{Wisely2020} 
In a follow-up paper, distinction between mild cognitive impairment and normal cognition based on only OCT-A images (AUROC = $0.625$) underperformed classification on image-derived quantitative data (AUROC = $0.960$). These results indicate that it can be difficult to extract meaningful features from retinal imaging, especially in a small dataset.
\newline \newline 
CNNs rely on large image datasets, which is not always available in the medical imaging domain. To overcome this, generative artificial intelligence (AI) could extend the training dataset by synthesizing image data. Several approaches to generative AI exist, among which Diffusion Probabilistic Models (DDPMs).\parencite{DDPM-original}
DDPMs have gained significant popularity in the field of medical imaging 
and demonstrate superior output diversity compared to as Generative Adversarial Networks and Variational Auto Encoders.\parencite{Kazerouni2023} However,
DDPMs are more prone to memorization.\parencite{BewareOfDddpm, Dhariwal2021} 
DDPMs generate synthetic images by iteratively removing noise from an initial image made of pure Gaussian noise.\parencite{DDPM}
Class-conditioning the DDPM allows one model to generate images with varying content, for example of a specific animal, or medical images of a specific disease.
\newline \newline
This study aims to predict AmyloidPET status with multimodal retinal imaging and to improve the performance through pretraining with synthetic data. A DDPM was developed to generate synthetic images for four types of retinal scans and a filter was created to recognize realistic synthetic images. Lastly, a multimodal classifier was trained to fuse unimodal predictions and patient information. The design of this framework serves as proof of concept for leveraging generative AI in classification tasks in ophthalmology. 

%% file: Methods/Datasets.tex
\subsection{Participants}
Data from 183 patients of two retrospective cohorts with ophthalmic examinations including FSLO, conventional OCT and OCT-A were used. 328 eyes of 167 subjects were used: 116 eyes of 59 AmyloidPET positive (AmyloidPET+) subjects and 212 eyes of 108 AmyloidPET negative (AmyloidPET-) subjects. 203 eyes were part of the PreclinAD cohort consisting of cognitively healthy participants (monozygotic twins) aged $\geq$ 60.\parencite{Boomsma2006,Konijnenberg2018,tenKate2018,vandeKreeke2018} The remaining 125 eyes were part of an ongoing trial of the Alzheimer Center Amsterdam (METC 2019.623) which included participants aged $\geq$ 50 years. See Appendix for further description of the cohorts. 
\subsection{Datasets}
The dataset contained in total 30 different modalities extracted from FSLO, OCT and OCT-A examinations. 
Four modalities from three scanner types covering the macula, optic nerve had (ONH) and fundus were selected for image synthesis and as inputs to the classification networks: 1) 2D OCT-A of the superficial retinal layers of the macula (OCTA-SMAC); 2D OCT B-Scan (OCT-B) of the 2) ONH (OCT-BONH); and 3) macula (OCT-BMAC); and 4) 2D FSLO autofluorescence (FAF). Images in the datasets were included based on image quality and the modalities were selected after the development of generative models for a larger set of eight modalities, as described in the Appendix. 
We created overlapping collections for the development of: generative models ($D_{synth}$); a modality recognition classifier ($D_{filter}$); unimodal classifiers ($D_{uni}$); and multimodal classifiers ($D_{multi}$). Of the $328$ eyes in this dataset, $198$ eyes were included in $D_{multi}$, $326$ in $D_{uni}$ and $328$ in $D_{synth}$ and $D_{filter}$ (Table \ref{tab:datasets}). 
 

\input{Tables/table_datasets_t}


%% file: Tables/table_datasets_t.tex
                                        
\begin{table}[H]
\footnotesize
\centering
\begin{tabular}{lllllllll}
\hline
Subset                                         & \begin{tabular}[c]{@{}l@{}}{[}N{]} eyes \\ (All)\end{tabular} & \begin{tabular}[c]{@{}l@{}}{[}N{]} Images \\ (All)\end{tabular} & \begin{tabular}[c]{@{}l@{}}{[}N{]} eyes \\ (Dev)\end{tabular} & \begin{tabular}[c]{@{}l@{}}{[}N{]} images \\ (Dev)\end{tabular} & \begin{tabular}[c]{@{}l@{}}{[}\%{]} AmyloidPET+ \\ (Dev)\end{tabular} & \begin{tabular}[c]{@{}l@{}}{[}N{]} eyes \\ (Test)\end{tabular} & \begin{tabular}[c]{@{}l@{}}{[}N{]} images \\ (Test)\end{tabular} & \begin{tabular}[c]{@{}l@{}}{[}\%{]} AmyloidPET+ \\ (Test)\end{tabular} \\ \hline
\multicolumn{1}{l|}{Multimodal classification} & 198                                                           & 198                                                             & 160                                                           & 160                                                             & 0.394                                                                 & 38                                                             & 38                                                               & 0.368                                                                  \\ \hline
\multicolumn{1}{l|}{Unimodal classification}   & 326                                                           & 1080                                                            & 285                                                           & 869                                                             &                                                                       & 83                                                             & 211                                                              &                                                                        \\
\multicolumn{1}{l|}{- OCTA-SMAC}               & 276                                                           & 276                                                             & 223                                                           & 223                                                             & 0.400                                                                   & 53                                                             & 53                                                               & 0.396                                                                  \\
\multicolumn{1}{l|}{- OCT-BONH}                & 278                                                           & 278                                                             & 220                                                           & 220                                                             & 0.429                                                                 & 58                                                             & 58                                                               & 0.379                                                                  \\
\multicolumn{1}{l|}{- OCT-BMAC}                & 276                                                           & 276                                                             & 220                                                           & 220                                                             & 0.426                                                                 & 56                                                             & 56                                                               & 0.411                                                                  \\
\multicolumn{1}{l|}{- FAF}                     & 250                                                           & 250                                                             & 206                                                           & 206                                                             & 0.354                                                                 & 44                                                             & 44                                                               & 0.364                                                                  \\ \hline
\multicolumn{1}{l|}{Synthesis}                 & 328                                                           &  1102                                                                &                                                               &                                                                 &                                                                       &                                                                &                                                                  &                                                                        \\
\multicolumn{1}{l|}{- OCTA-SMAC}               & 276                                                           & 276                                                             & 223                                                           & 223                                                             & 0.386                                                                 & 53                                                             & 53                                                               & 0.396                                                                  \\
\multicolumn{1}{l|}{- OCT-BONH}                & 278                                                           & 278                                                             & 220                                                           & 220                                                             & 0.377                                                                 & 58                                                             & 58                                                               & 0.379                                                                  \\
\multicolumn{1}{l|}{- OCT-BMAC}                & 276                                                           & 276                                                             & 220                                                           & 220                                                             & 0.382                                                                 & 56                                                             & 56                                                               & 0.411                                                                  \\
\multicolumn{1}{l|}{- FAF}                     & 254                                                           & 272                                                             & 210                                                           & 225                                                             & 0.351                                                                 & 44                                                             & 47                                                               & 0.340                                                                   \\ \hline
\end{tabular}
\caption{Datasets used for training and validation (development, Dev) and testing (Test). Images without known AmyloidPET status were excluded from the classification sets. The first split between testing, training and validation was created for $D_{multi}$ with $198$ eyes with known AmyloidPET status and available images for all four modalities. The remaining $128$ eyes with images for at least one of the selected modalities and known AmyloidPET status were distributed over these splits to create the $D_{uni}$ collection. This totals to $1080$ images used for unimodal classification. $22$ images with known AmyloidPET status from duplicate FAF recordings were added to these splits to create $D_{synth}$. The training, validation and test splits of the collections were created in this way to prevent information leakage between training and evaluation sets of the collections.}
\label{tab:datasets}
\end{table}

%% file: Methods/Models.tex
\subsection{Models}
The pipeline for data synthesis and for training the classification networks are depicted in Figure \ref{fig:models_overview}. Our approach involved the generation of a synthetic image dataset, development of a filter to ensure high-quality synthetic images, and training unimodal and multimodal classifiers to predict AmyloidPET status. See Appendix for details of the network architectures and hyperparameters governing the training trajectories.
\input{Figures/pipeline_t}

\subsubsection{Synthetic images}
We used a conditional U-Net DDPM for generating synthetic images corresponding to specific AmyloidPET status.
The DDPM model was selected based on the quality of the synthetic images and the required training time. 
A potential for memorization of the training images exists in a dataset with DDPMs.
Therefore, we evaluated the diversity and uniqueness of generated images 
using a random sample of $200$ synthetic images and all real images.\parencite{BewareOfDddpm} The maximum Pearson's correlation coefficient (pearsonr) of one image from an arbitrary set $A$ with all images in a different set $B$ expresses how similar this image is at most to set $B$. In the case of memorization the distribution of maximum correlation values of all synthetic images with the set of real images would be high. The diversity of synthetic images was evaluated by comparing the distribution of maximum pearsonr values among real images and among synthetic images. With Wasserstein Distance (WD) we expressed the difference between two distributions of correlation values, Kolmogorov–Smirnov (KS) test informed the significance of such differences. \newline

\subsubsection{Filter}
We assumed that for generated images with large artifacts or deformations, it would be more difficult to recognize the modality type. Therefore, a CNN was trained to recognize the modality of images and then applied to detect unrealistic synthetic images. Images with incorrectly recognized modalities were discarded. The filter was evaluated with Matthew's Correlation Coefficient (MCC) for a balanced evaluation of the filter's overall performance by correlating the outputs, between 0 and 1 for each class, and the ground truth labels.\parencite{Chicco2020}
\newline
\subsubsection{Classification}
Unimodal classifiers were developed for all modalities to predict AmyloidPET status. 
The EfficientNet-B0 backbone without pretrained ImageNet weights showed best loss reduction in small (25 trials) hyperparameter optimization experiments.\parencite{EfficientNet} Unimodal classifiers were pretrained with synthetic images and subsequently finetuned on real data. 
The FC of multimodal classifiers was trained with outputs of the unimodal classifiers, per eye, and metadata.
Performance was evaluated on the validation and test sets, reported by area under the receiver operator curve (AUROC), area under the precision recall curve (AUPR), sensitivity, specificity and F1-score as harmonic mean between sensitivity and precision. 
With class activation maps (CAMs) we attempted to gain insights into the CNN predictions. CAMs indicated which discriminative image regions contributed to a model's output value.\parencite{GradCAM}
As our model outputted the probability for AmyloidPET-, the resulting heatmap displayed the image regions that contributed to an output of a higher probability of AmyloidPET-. 

%% file: Figures/pipeline_t.tex
\begin{figure}[H]
\centering
\includegraphics[width=\linewidth]{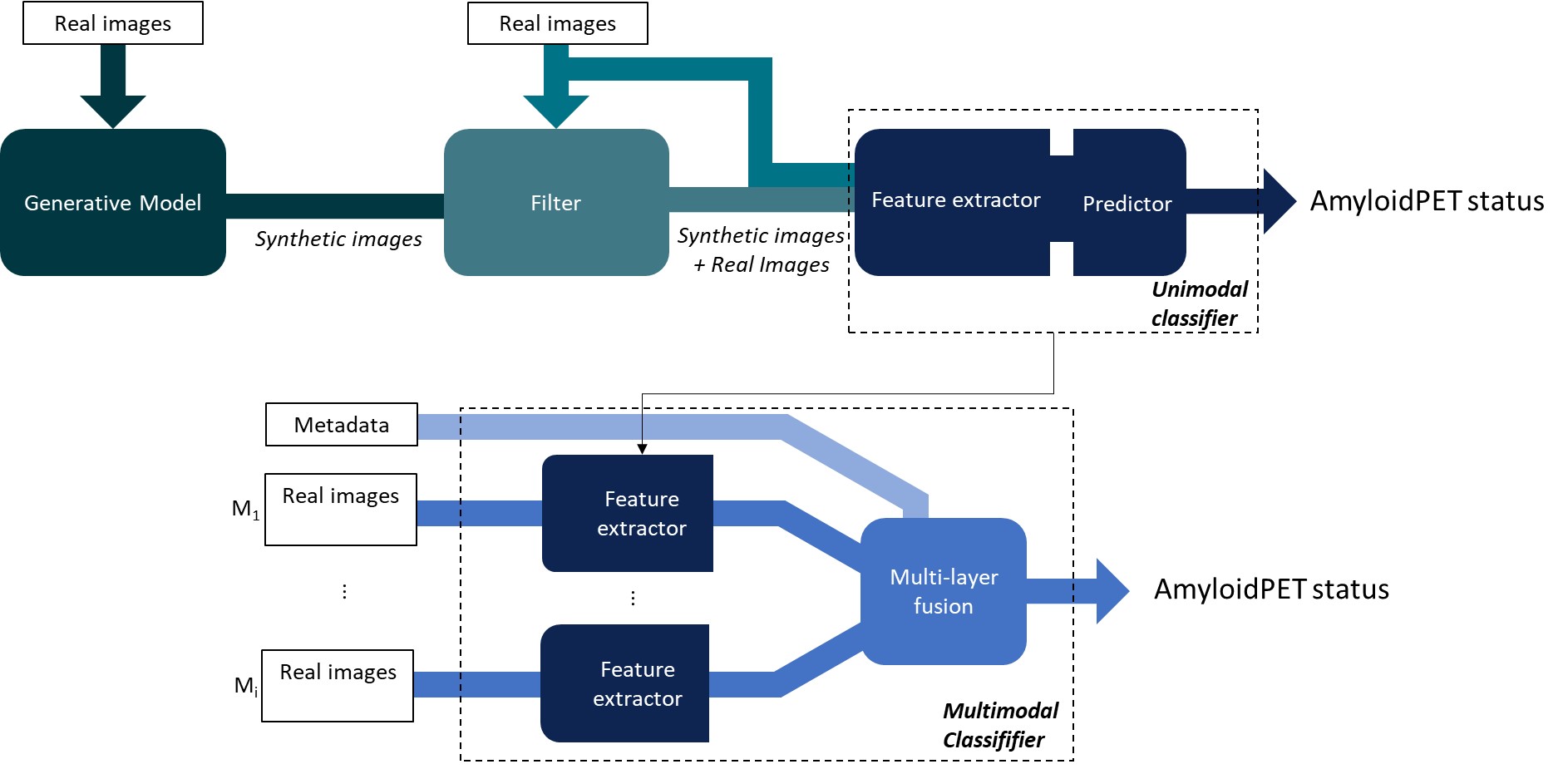}
\caption{Illustration of the pipeline. (Top): Synthetic images were generated by a DDPM. The synthetic images for which the filter could recognize the modality were included in the training budget of 1000 synthetic images per class. Both synthetic and real images were used to train unimodal classifiers for predicting AmyloidPET status. We created 'baseline' unimodal classifiers trained on real images, and 'pretrained' unimodal classifiers pretrained on $1000$ synthetic images per class and finetuned on real images. Unimodal classifiers were not trained with metadata inputs because synthetic data, which has no associated age and gender, was used for pretraining. (Bottom): We compared multimodal classifiers with the baseline and pretrained unimodal classifiers. The weights of unimodal classifiers were fixed after training. A three-layer fully connected network (FC) performed late heterogeneous fusion of the unimodal predictions and metadata into one AmyloidPET probability prediction. If metadata was included as inputs, age (binary) and gender (scaled by $0.01$) metadata were also fed to the FC. Output of the unimodal and multimodal classifiers were scored between 0 and 1 for the probability of AmyloidPET negative status.
}
\label{fig:models_overview}
\end{figure}

%% file: Results/Synthesis.tex
\subsection{Synthetic data}
Examples of synthetic images are shown in Figure \ref{fig:synthetic_images} in the Appendix. Figure \ref{fig:best_3_correlations} displays, for each modality, the three synthetic images with the highest correlation with any real image and the corresponding real image. Visual inspection of these examples illustrates that it depended on the nature of a modality whether pearsonr of the pixel values was in good agreement with the visual similarity of retinal images.
Figure \ref{fig:correlations_subfigs} displays distributions of the maximum correlation for synthetic images with all real images (SvR), among the real images (RvR), and among synthetic images (SvS). We observed little memorization in the generated data as the SvR distributions did not reach close to 1. However, we discoverd a positive trend between SvR and RvR values (pearsonr =$0.997$): if for a modality the similarity amongst real images was high, the similarity between real images and synthetic images was also high. This could be caused by characteristics of the modalities (Figure \ref{fig:best_3_correlations}). We compared SvR with RvR for each modality to provide extra context on the extend of memorization. Similarity of synthetic images to real images did not exceed the similarity among real images for any of the modalities, which weakens the concern for memorization. SvS was stronger than RvR for OCTA-SMAC (WD = $0.025$, p=$1.22$e$^{-6}$) and FAF (WD = $0.027$, p=$8.079$e$^{-11}$), implying reduced diversity of the synthetic images of these modalities. However, the distribution plots in Figure \ref{fig:correlations_subfigs} display that these differences between between the RvR and SvS distributions are minimal.
\input{Figures/best_3_correlations_t}
\input{Figures/correlations_subfigs_t}

\subsection{Filter}
The trained filter model achieved $99\%$ accuracy for recognizing the modalities on the test set with MCC of $0.990$ for the predictions on the validation set and MCC of $0.997$ on the test set (Table \ref{tab:filter_accuracy}). The model performed so well that it could correctly recognize the modality of unrealistic synthetic images of type OCTA-SMAC, OCT-SONH and OCT-SMAC, diminishing its role in detecting unrealistic images.
Therefore, we applied a manually set threshold on the model outputs for these three modalities; $0.90$, $0.99$ and $0.96$, respectively. Any sampled synthetic image was included in our synthetic dataset if the predicted modality was correct and if the confidence for this modality satisfied this threshold.


%

%% file: Figures/best_3_correlations_t.tex
\begin{figure}[H]
     \centering
     \begin{subfigure}[b]{0.4\textwidth}
         \centering
         \includegraphics[width=\textwidth]{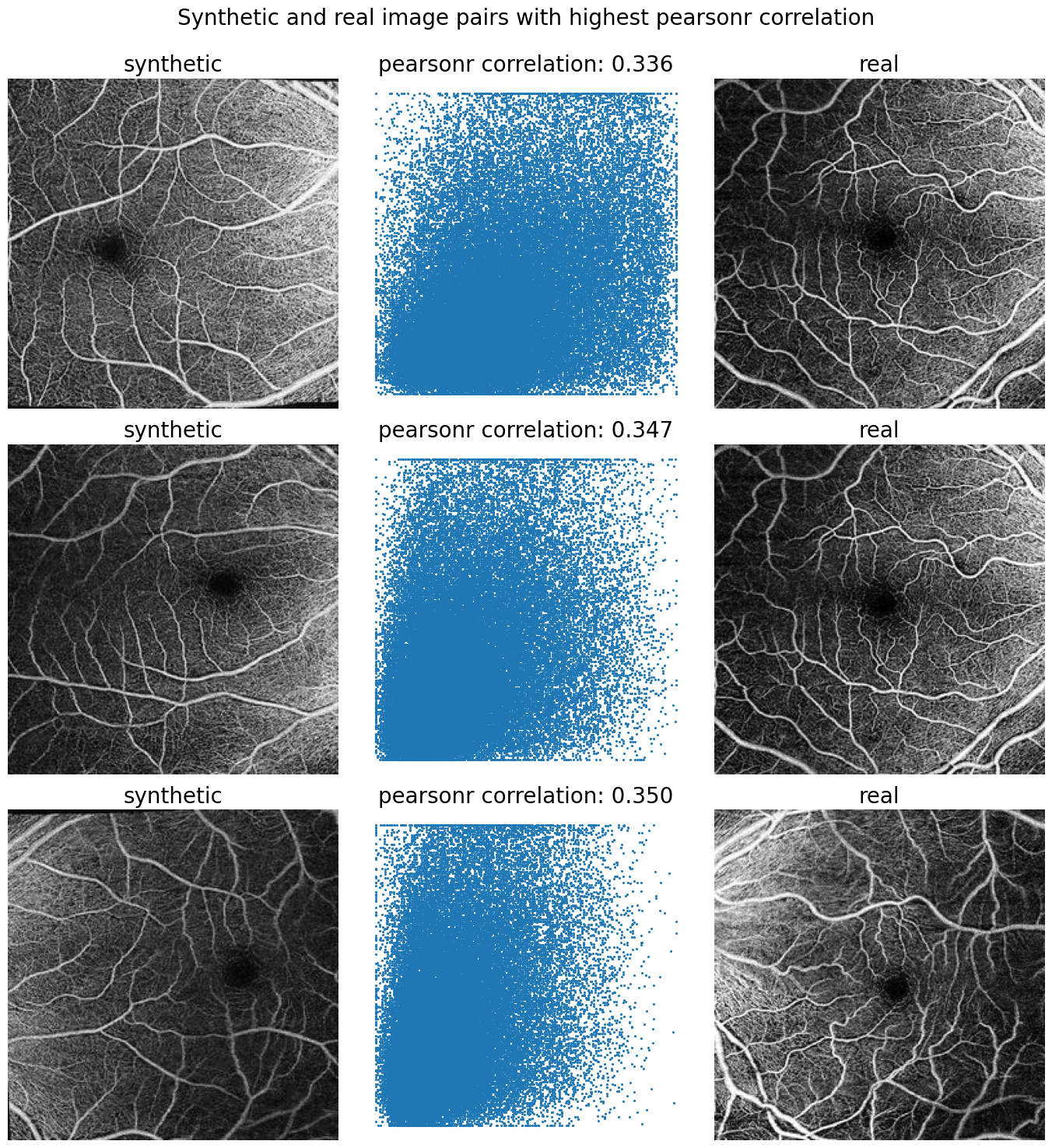}
         \caption{ }
     \end{subfigure}
     \hfill
      \begin{subfigure}[b]{0.4\textwidth}
         \centering
         \includegraphics[width=\textwidth]{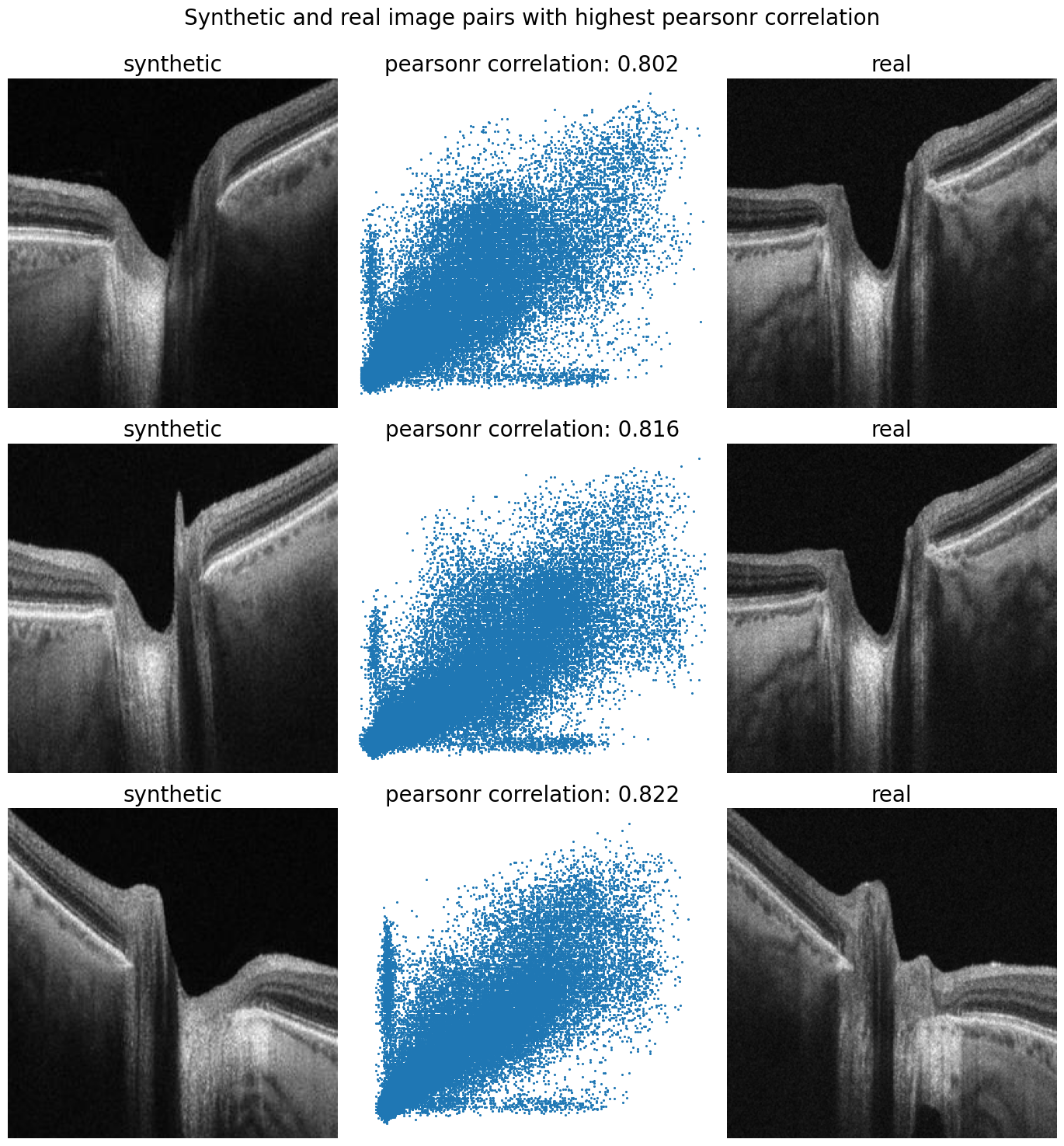}
         \caption{ }
     \end{subfigure}
     \vfill
     \begin{subfigure}[b]{0.4\textwidth}
         \centering
         \includegraphics[width=\textwidth]{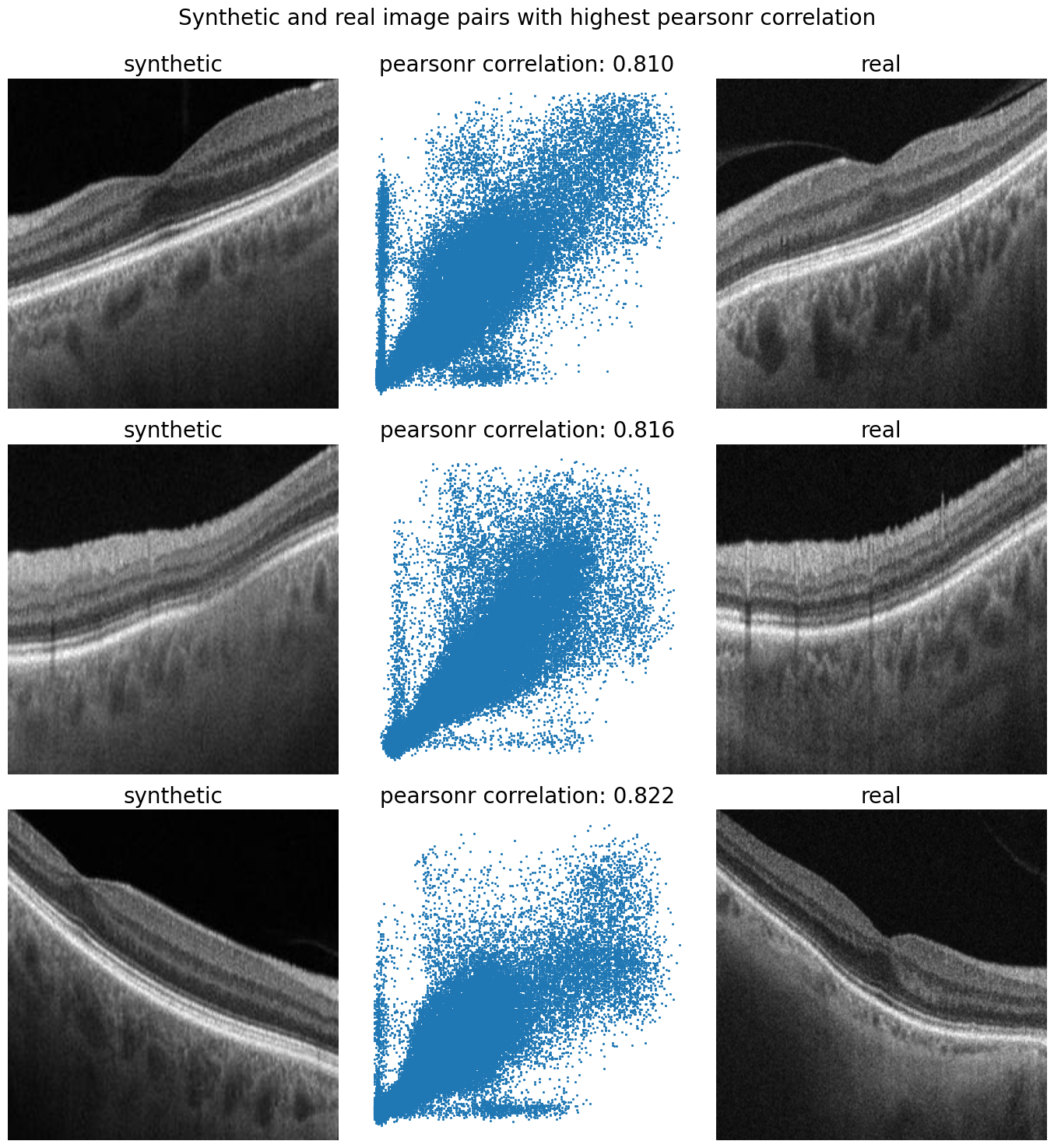}
         \caption{ }
     \end{subfigure}
     \hfill
     \begin{subfigure}[b]{0.4\textwidth}
         \centering
         \includegraphics[width=\textwidth]
         {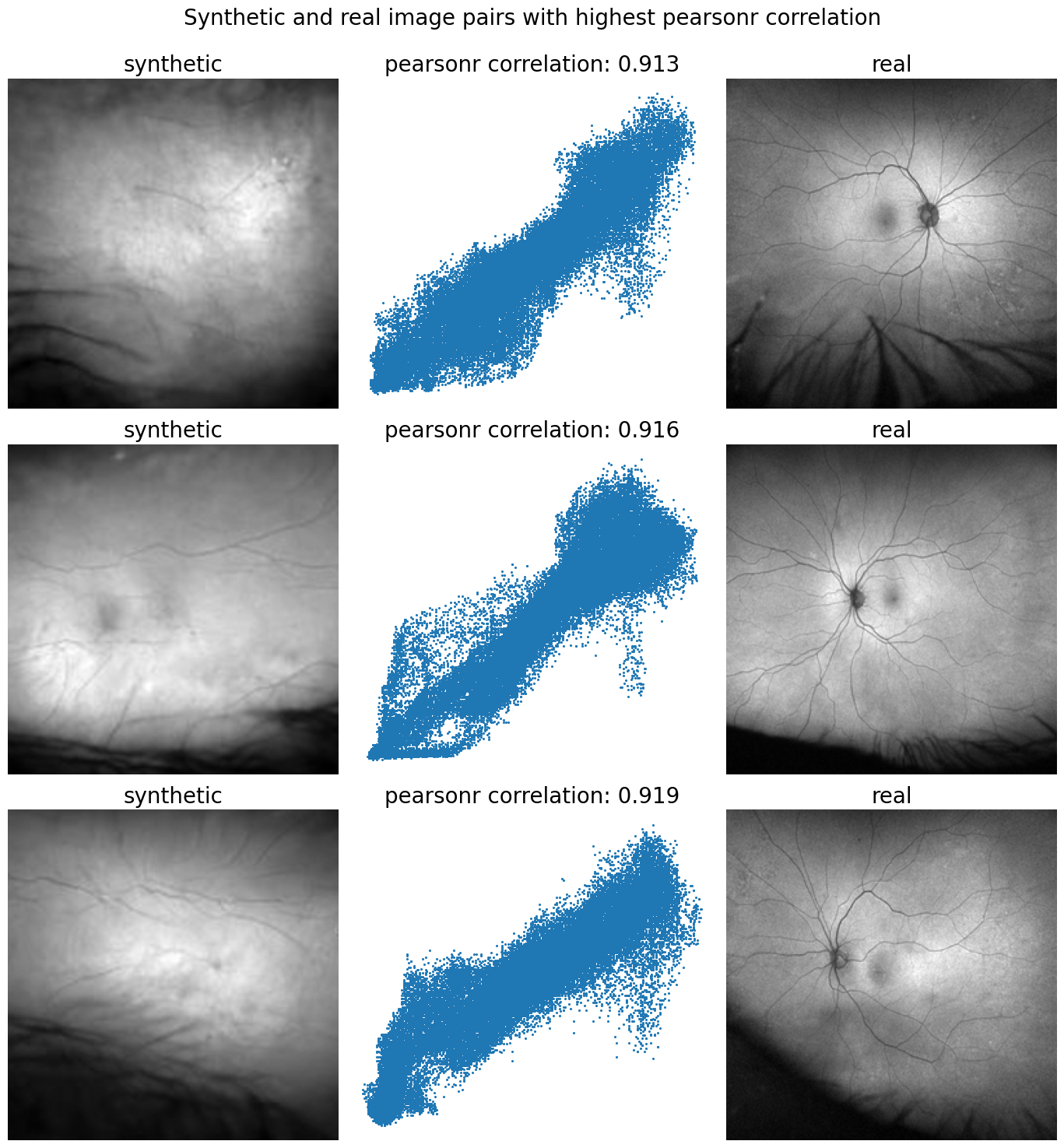}
         \caption{ }
     \end{subfigure}
    \caption{Examples of the synthetic images with the highest correlation to any real image. Pairs of synthetic images and the corresponding real image that it most closely resembles are displayed together with scatter plot of the pixel values and the correlation value. (a) OCTA-SMAC; (b) OCT-BONH; (c) OCT-BMAC; (d) FAF. For OCT-BMAC and OCT-BONH the synthetic images strongly resembled the real images but were not exact copies. For OCTA-SMAC and FAF the images with the highest correlations showed less resemblance. This was also reflected by the lower correlation values for OCTA-SMAC. FAF exhibited the highest distribution of maximum correlation values. These images consisted of a predominantly grey background which contributed to a high correlation between any two images of this modality. }
    \label{fig:best_3_correlations}
\end{figure}

%% file: Figures/correlations_subfigs_t.tex
\begin{figure}[H]
     \centering
     \begin{subfigure}[b]{0.49\textwidth}
         \centering
         \includegraphics[width=\textwidth]{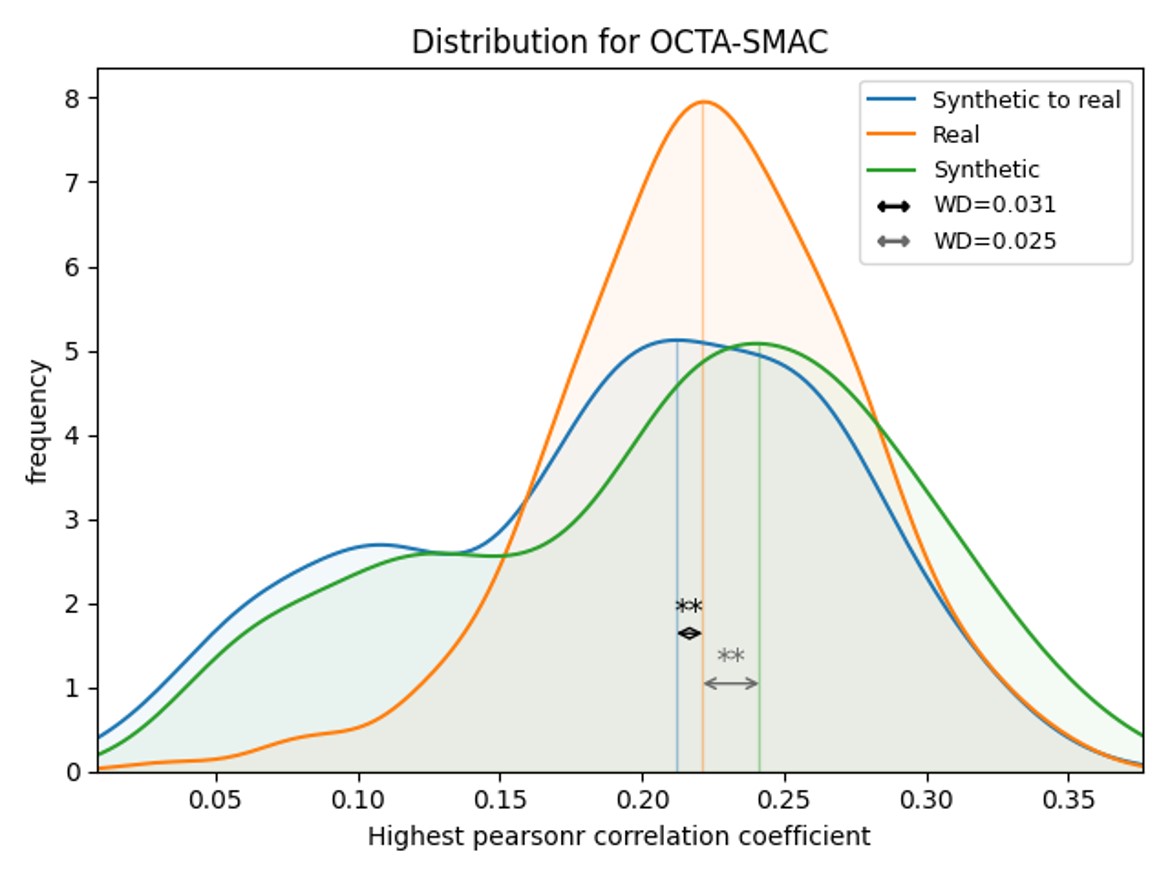}
         \caption{ }
     \end{subfigure}
      \begin{subfigure}[b]{0.49\textwidth}
         \centering
         \includegraphics[width=\textwidth]{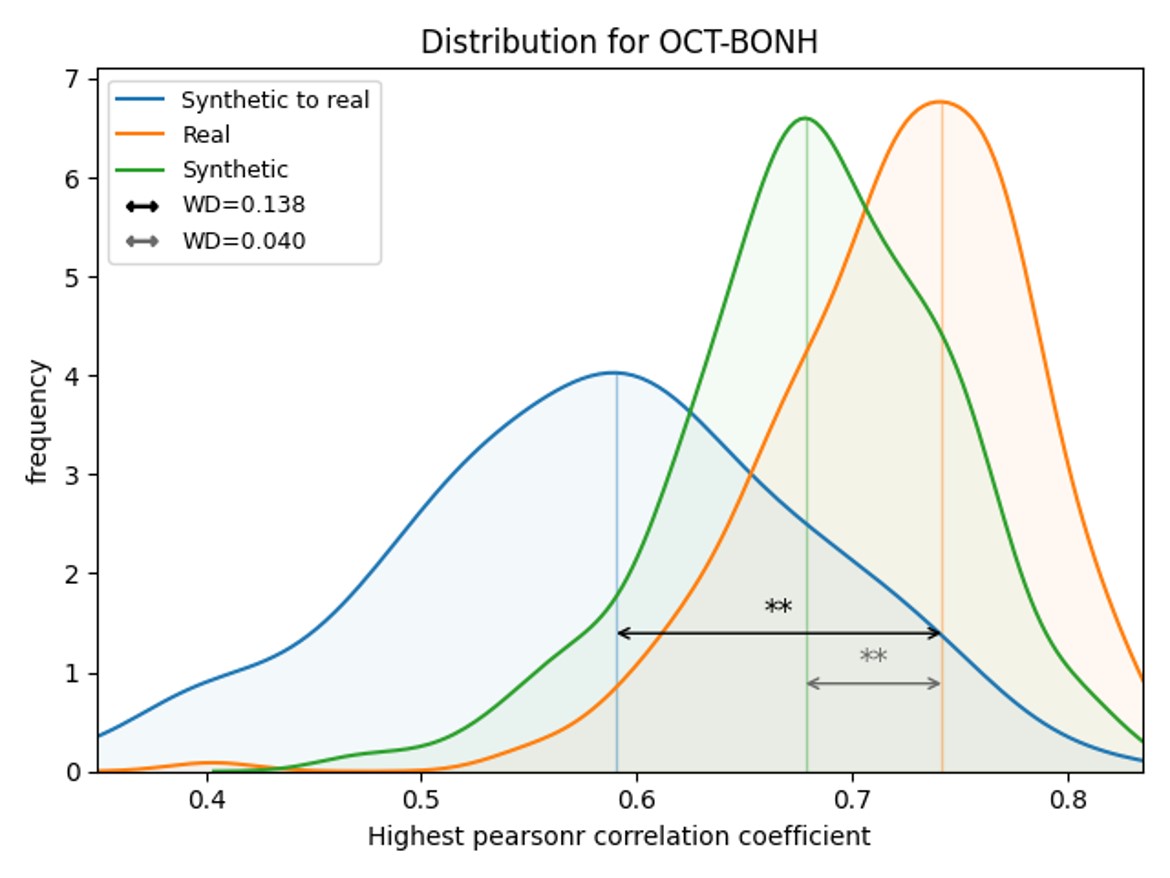}
         \caption{ }
     \end{subfigure}
     \begin{subfigure}[b]{0.49\textwidth}
         \centering
         \includegraphics[width=\textwidth]{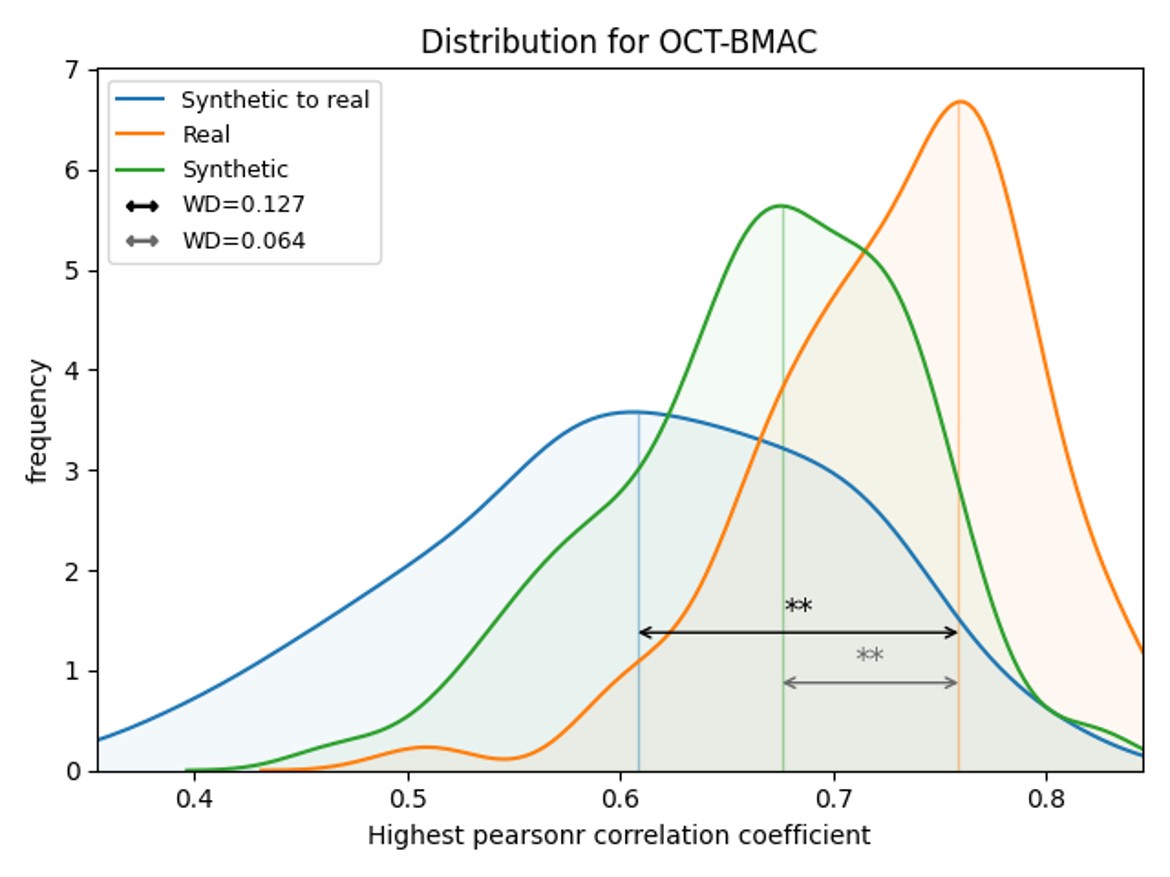}
         \caption{ }
     \end{subfigure}
     \begin{subfigure}[b]{0.49\textwidth}
         \centering
         \includegraphics[width=\textwidth]{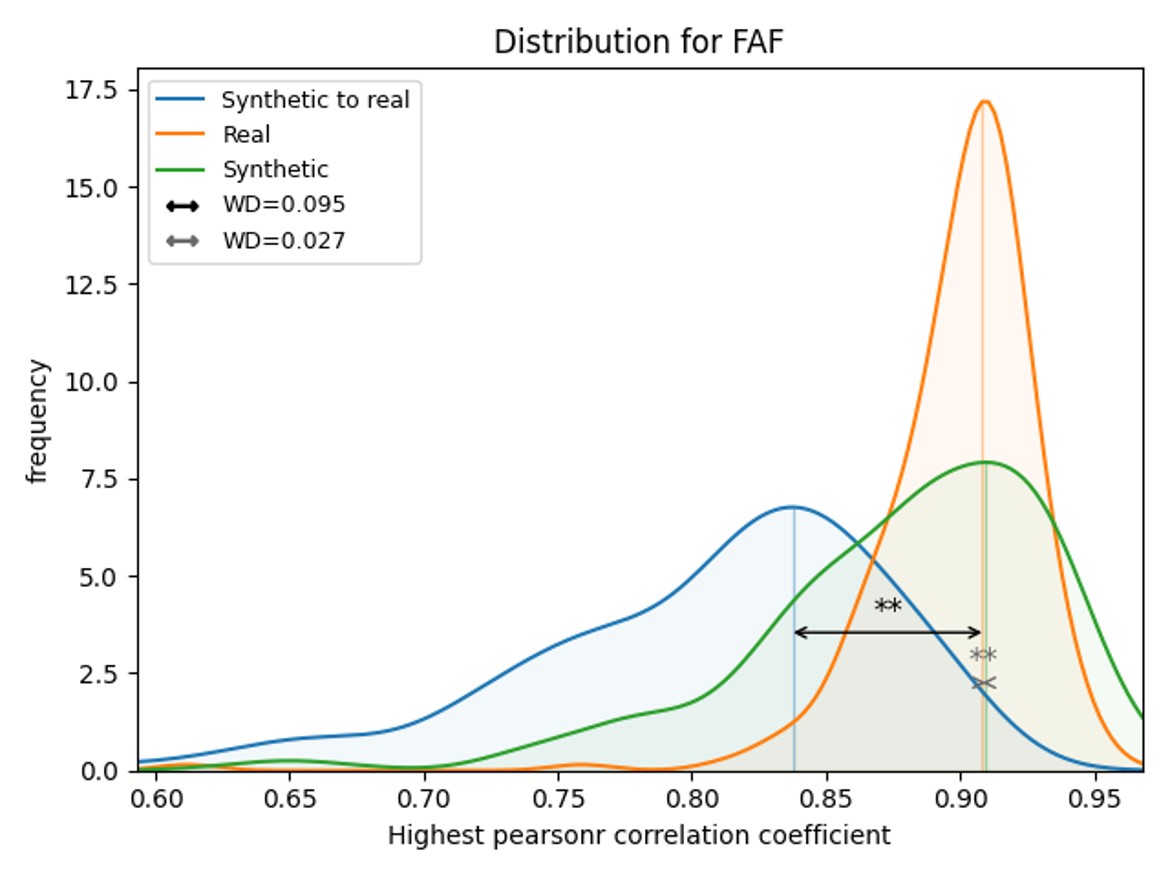}
         \caption{ }
     \end{subfigure}
    \caption{(A-D): Distributions for maximum pearsonr values computed for $200$ synthetic images and all real images. Distributions display the highest correlation for image pairs among real images (RvR, orange), among synthetic images (SvS, green) and for all synthetic images with any real image (SvR, blue). Arrows indicate the differences between SvR and RvR distributions (black) and the differences between RvR and SvS distributions (grey). WD values express the distance between two distributions with KS test p-values for the significance of such differences. pearsonr = Pearson's correlation coefficient. ** = p < 0.005}
    \label{fig:correlations_subfigs}
\end{figure}

%% file: Results/Classification.tex
\subsection{Classification}
Table \ref{tab:classification_results_2} displays the results of classification experiments. 
Pretraining on synthetic data showed slight AUPR improvement for OCT-BONH, OCT-BMAC and FAF on the validation set, as well as for OCTA-SMAC and OCT-BMAC on the test set. 
The overall best model on the test set was the multimodal classifier trained on real data (AUPR $0.634$, F1-score $0.625$). The best unimodal classifier in terms of AUPR was OCTA-SMAC trained on synthetic data (AUPR $0.613$). In terms of F1-score, the pretrained OCT-BMAC classifier performed best (F1-score $0.596$).
In several models, finetuning on real data showed a reduction of performance compared to the models trained on synthetic data.
\input{Tables/classfication_results_2_t}


%% file: Tables/classfication_results_2_t.tex
\begin{table}[]
\centering
\scriptsize
\begin{tabular}{l|lll|lll|llllll|lll}
                & \multicolumn{3}{c|}{AUPR}                        & \multicolumn{3}{c|}{AUROC}                       & \multicolumn{3}{c}{F1-score}                                          & \multicolumn{3}{c|}{Sensitivity}                 & \multicolumn{3}{c}{Specificity}                  \\
                & Real           & Synth.     & Pretr.         & Real           & Synth.      & Pretr.         & Real           & Synth.      & \multicolumn{1}{l|}{Pretr.}         & Real           & Synth.      & Pretr.         & Real           & Synth.      & Pretr.         \\ \hline
{\ul Validation}       &                &                &                &                &                &                &                &                & \multicolumn{1}{l|}{}               &                &                &                &                &                &                \\
Unimodal        &                &                &                &                &                &                &                &                & \multicolumn{1}{l|}{}               &                &                &                &                &                &                \\
- OCTA-SMAC     & 0.583          & \textbf{0.632} & 0.436          & 0.646          & \textbf{0.690} & 0.583          & 0.635          & \textbf{0.695} & \multicolumn{1}{l|}{0.565}          & \textbf{0.909} & 0.740          & 0.591          & 0.364          & 0.610          & \textbf{0.667} \\
- OCT-BONH      & 0.573          & 0.593          & \textbf{0.617} & 0.602          & 0.617          & \textbf{0.707} & 0.553          & 0.560          & \multicolumn{1}{l|}{\textbf{0.690}} & 0.542          & 0.510          & \textbf{0.833} & 0.688          & \textbf{0.690} & 0.563          \\
- OCT-BMAC      & 0.611          & 0.569          & \textbf{0.665} & 0.668          & 0.577          & \textbf{0.728} & 0.655          & 0.605          & \multicolumn{1}{l|}{\textbf{0.667}} & \textbf{0.826} & 0.650          & 0.739          & 0.484          & 0.500          & \textbf{0.645} \\
- FAF           & 0.545          & \textbf{0.864} & 0.576          & 0.744          & \textbf{0.858} & 0.664          & 0.629          & \textbf{0.811} & \multicolumn{1}{l|}{0.519}          & 0.647          & \textbf{0.835} & 0.412          & 0.774          & 0.775          & \textbf{0.903} \\
Multimodal      &                &                &                &                &                &                &                &                & \multicolumn{1}{l|}{}               &                &                &                &                &                &                \\
- No metadata   & 0.406          & -              & \textbf{0.632} & 0.567          & -              & \textbf{0.708} & 0.600          & -              & \multicolumn{1}{l|}{0.571}          & \textbf{0.800} & -              & 0.533          & 0.458          & -              & \textbf{0.792} \\
- With metadata & 0.456          & -              & \textbf{0.467} & \textbf{0.631} & -              & 0.592          & 0.588          & -              & \multicolumn{1}{l|}{0.636}          & 0.667          & -              & \textbf{0.933} & \textbf{0.625} & -              & 0.375          \\
{\ul Test}      &                &                &                &                &                &                &                &                & \multicolumn{1}{l|}{}               &                &                &                &                &                &                \\
Unimodal        &                &                &                &                &                &                &                &                & \multicolumn{1}{l|}{}               &                &                &                &                &                &                \\
- OCTA-SMAC     & 0.338          & \textbf{0.613} & 0.455          & 0.381          & \textbf{0.647} & 0.586          & 0.381          & \textbf{0.541} & \multicolumn{1}{l|}{0.522}          & 0.381          & 0.476          & \textbf{0.571} & 0.594          & \textbf{0.813} & 0.594          \\
- OCT-BONH      & \textbf{0.488} & 0.368          & 0.481          & 0.530          & 0.412          & \textbf{0.583} & 0.533          & \textbf{0.545} & \multicolumn{1}{l|}{0.489}          & 0.727          & \textbf{0.955} & 0.500          & 0.389          & 0.056          & \textbf{0.667} \\
- OCT-BMAC      & 0.350          & 0.451          & \textbf{0.579} & 0.391          & \textbf{0.569} & 0.615          & 0.426          & 0.571          & \multicolumn{1}{l|}{\textbf{0.596}} & 0.435          & 0.696          & \textbf{0.739} & \textbf{0.576} & 0.485          & 0.485          \\
- FAF           & \textbf{0.390} & 0.310          & 0.347          & \textbf{0.549} & 0.326          & 0.478          & \textbf{0.571} & 0.100          & \multicolumn{1}{l|}{0.489}          & \textbf{0.875} & 0.063          & 0.688          & 0.321          & \textbf{0.893} & 0.357          \\
Multimodal      &                &                &                &                &                &                &                &                & \multicolumn{1}{l|}{}               &                &                &                &                &                &                \\
- No metadata   & 0.486          & -              & \textbf{0.441} & \textbf{0.622} & -              & 0.491          & \textbf{0.500} & -              & \multicolumn{1}{l|}{0.333}          & \textbf{0.500} & -              & 0.286          & 0.708          & -              & \textbf{0.750} \\
- With metadata & \textbf{0.634} & -              & 0.306          & \textbf{0.729} & -              & 0.369          & \textbf{0.625} & -              & \multicolumn{1}{l|}{0.531}          & 0.714          & -              & \textbf{0.929} & \textbf{0.667} & -              & 0.083         
\end{tabular}
\caption{Classification results for the validation set and test set. Real = trained on real data. Synth. = trained on synthetic data. Pretr = pretrained on synthetic data and finetuned on real data. Precision and recall were determined based on a threshold obtained with the youden's index.}
\label{tab:classification_results_2}
\end{table}

%% file: Results/Cams.tex
\subsection{Class activation maps}
Figures \ref{fig:cams_real_classifiers}-\ref{fig:cams_synthetic_classifiers} display test images with GradCAMs that identified the regions of the image that contributed to a higher model output. 
We compared the GradCAMS in Figure \ref{fig:cams_real_classifiers} and \ref{fig:cams_pretrained_classifiers} to interpret and compare the outputs of the baseline and pretrained classifiers. We also reviewed the GradCAMS of the classifier trained on synthetic images (Figure \ref{fig:cams_synthetic_classifiers}) to discover whether a model trained on synthetic data could identify relevant areas in the images. We observed different responses to real images by the various classifiers. For example, in OCTA-SMAC, attention of the baseline classifier to blood vessels in the periphery switched to the center of the image by the pretrained classifier. The baseline OCT-BONH classifier showed high response to small areas in the layers of the retina. The shape and location of the attention in the OCT-BONH models implied that meaningful features were learned, however the output values for the different classes were very close to each other ($0.494$ and $0.485$ compared to $0.471$ and $0.489$) which implied that the classes were not well distinguished by these features. The pretrained FAF classifier identified areas around the fovea and ONH.

%% file: Discussion/Discussion.tex
To our knowledge, this is the first study to generate multimodal synthetic image data to detect AmyloidPET status using retinal imaging. \citeauthor{BewareOfDddpm}\parencite{BewareOfDddpm} noted that diffusion models are prone to memorizing training images, especially with small datasets. However, we were able to construct a DDPM that was capable of generating synthetic retinal images with our small dataset. The synthetic images were unique and not copies of the limited number of real images we provided. Furthermore, we hypothesized that these synthetic images contained relevant information for predicting AmyloidPET status on real images. A small dataset can lead to a classifier’s inability to recognize salient image features due to overfitting on the training data. To mitigate this, we supplemented the development set with 1000 images per AmyloidPET class, generated by a conditional DDPM. Pretraining CNNs with synthetic data slightly improved classification performance for two out of four modalities in terms of AUPR and improved F1-score for one modality. This suggests that exploiting synthetic data can enhance CNN performance in small datasets for medical image classification. Further research into synthetic data training budgets, synthesis methods, and exploitation strategies could demonstrate the potential of generative AI for deep learning in medical imaging.
\newline \newline
Our best unimodal and multimodal classifiers were not pretrained on synthetic data. Finetuning on real data sometimes reduced performance of unimodal classifiers, possibly due to low similarity of the real and synthetic images of these modalities. Our best model used multimodal FSLO, OCT, and OCT-A inputs and metadata, and achieved AUPR of $0.634$ (AUROC $0.729$) on the test set, outperforming our best unimodal classifiers. 
By including metadata, the performance of the multimodal classifier improved from AUPR $0.486$ (AUROC $0.622$). \citeauthor{Wisely2020}\parencite{Wisely2020} and \citeauthor{Cheung2022}\parencite{Cheung2022} observed similar improvements, for AD and AmyloidPET prediction respecitively, although the differences were not significant. 
It should be noted that AmyloidPET detection differs from AD diagnosis as cognitively healthy individuals can have a positive AmyloidPET scan. Additionally, it may be worthwhile to investigate the effect of unilateral and bilateral inputs for our models as \citeauthor{Cheung2022}\parencite{Cheung2022} found that bilateral inputs improved AmyloidPET status prediction compared to unilateral inputs (AUROC = $0.68$ - $0.86$ vs. $0.61$ - $0.83$ on external validation). However, it must be noted that we would have limited training samples for the bilateral predictions due to our dataset size.
\newline \newline
In our study, FAF-based models performed poorly compared to other classifiers, aligning with the results of \citeauthor{Wisely2020}\parencite{Wisely2020}, as they concluded that FSLO images have low utility for predicting AD diagnosis. Our suggestion for future experiments is to explore metadata incorporation into unimodal classifiers, which requires adaptation for pretraining without metadata and subsequent finetuning with metadata. However, incorporating metadata can have mixed effects. \citeauthor{Cheung2022}\parencite{Cheung2022} argue that not requiring patient data is advantageous. Although excluding patient information such as age and gender could prevent CNNs from biasing towards patient groups of certain demographics, learning from patient age and gender may actually be valuable as these characteristics can affect retinal structure, aiding feature extraction and prediction.\parencite{Polascik2020,Munk2021}
The superior performance with metadata observed by \citeauthor{Wisely2020}\parencite{Wisely2020} as well as our study may be due to the inclusion of additional OCT-derived modalities alongside FSLO, which provide higher axial resolution. This diversity of inputs can offer more opportunity for utilization of metadata in processing unimodal predictions.
\newline \newline
As the first study to use synthetic retinal images for AmyloidPET prediction, we evaluated the models’ internal states with heatmaps to visualize learned knowledge. GradCAM heatmaps in Figures \ref{fig:cams_synthetic_classifiers} indicate that training on synthetic images can teach a model to recognize salient regions in real images. Specifically, a CNN based on OCTA-SMAC pretrained on synthetic images and finetuned on real images draws attention towards the center of the images, which is where the foveal avascular zone is located, an area associated with AD in several meta-analyses.\parencite{Jin2021,Ashraf2023}
Existing studies have provided saliency maps for fundus-based AD prediction and showed that small blood vessels and the main vascular branches are most salient.\parencite{Tian2021, Cheung2022} Our FAF classifier trained on real images projected similar attention on synthetic images, highlighting main vascular branches, with the synthetic-trained classifier showing similar attention on small vascular branches in real images.
\newline \newline
Limitations of this study are related to the nature of the training and validation sets. It is difficult to fit models that generalize well on unseen data because of our small dataset. Furthermore, performance on the evaluation dataset may not be good indicators of the performance of our methods, as the composition of the evaluation dataset influences the performance and may greatly vary depending on the split. We used stratified splits to address this but additional cross-validation experiments would yield more reliable results. A larger dataset could also mitigate these issues.





%% file: Supplementary/Dataset.tex
\subsection*{Cohorts}
203 eyes are part of the PreclinAD cohort which is an extension of the Amsterdam sub‐study of the European Medical Information Framework for AD from the Amsterdam UMC, location VUmc.\parencite{Konijnenberg2018,vandeKreeke2019} The PreclinAD cohort consists of cognitively healthy participants (monozygotic twins) from the Netherlands Twin Register who all underwent ophthalmic evaluation. The remaining 125 eyes are part of an ongoing trial of the Amsterdam Alzheimer Center (METC 2019.623). Both studies followed the Tenets of the Declaration of Helsinki and the Medical Ethics Committee of the VU University Medical Center in Amsterdam approved the studies. All participants signed an informed consent and underwent a screening protocol that included in short: ophthalmological examinations, (medical) history check-up, Mini Mental State Examination test and neuropsychological evaluations. Subjects with ischemic stroke, neurodegenerative disorders or systemic chronic conditions (i.e. Parkinson’s disease, Diabetes Mellitus andmultiple sclerosis) were excluded.

\subsection*{Dataset}
All scans were evaluated for image quality and availability of AmyloidPET status. Image quality for FSLO exams was reviewed based on movement or optical distortions such as lash or eyelid coverage and a straight eye gaze was required. OCT-A images were evaluated by focus and resolution and 3D OCT scans were evaluated on the same criteria. \newline \newline
The full image dataset covered retinal scans from three types of retinal modalities: Fundus SLO (Optos), OCT-A (Zeiss Angioplex) and volumetric OCT (Heidelberg OCT). The initial dataset included a set of 26 modalities. Optos: FAF and red-green fundus (FRG); Zeiss Angioplex: depth-encoded, and five layer-specific OCT-A and structural B-scan images for both the ONH and macula; Heidelberg: 3D volumetric cubes of the optic nerve and macula. Initial experiments were performed to create synthetic images for FAF, FRG, depth-encoded macula angiography (OCTA-EMAC), depth-encoded ONH angiography (OCTA-EONH), layer-specific OCT-A of the deep and superficial layers of the ONH (OCTA-DONH, OCTA-SONH) and macula (OCTA-DMAC, OCTA-SMAC). We also experimented with 3D DDPM for volumetric scans of the ONH (OCT-VONH) and macula (OCT-VMAC). 
The experiments failed to create realistic volumetric, depth-encoded and FRG images. Therefore, we made a selection of four modalities extracted from FSLO, OCT-A and structural OCT scans to maintain the variety of the inputs to the classification models. For FSLO we selected the FAF modality as we could not create realistic synthetic images for FRG. For OCT-A we used the superficial layer of the macula instead of the depth-encoded OCTA-EONH and OCTA-EMAC as most research suggests that changes in foveal avascular zone and vessel density in superficial macula may correlate with AD progression.\parencite{Katsimpris2021,Jin2021,Ashraf2023} Instead of volumetric structural OCT we used 2D structural OCT-BONH and OCT-BMAC that were generated during the angiography recordings on the Zeiss Angioplex.
A comparison of the 2D real and synthetic images for the initial and final DDPMs is depicted in Supplementary Figure \ref{fig:synthetic_images}.

\subsection*{Data splits}
The first split between testing, training and validation was created for $D_{multi}$ with eyes with known AmyloidPET status and available images for all selected modalities. The remaining eyes with images of at least one of the selected modalities and known AmyloidPET status were distributed over these splits to create the $D_{uni}$ collection. Images with known AmyloidPET status from duplicate FAF recordings were added to these splits to create $D_{synth}$. Finally, the splits in $D_{synth}$ were supplemented with images of modalities not selected for classification as well as images of eyes with unknown AmyloidPET status to form $D_{filter}$. 
The number of eyes in test splits was not expanded after $D_{uni}$ was formed, meaning that supplementary eyes were added to the training and validation splits of $D_{synth}$ and $D_{filter}$.
We aimed to keep the proportion of AmyloidPET positives (AmyloidPET+) balanced while assigning $20\%$ of the eyes to the test set and using $20\%$ of the eyes in a collection for validation. Splits for these sets were performed at a family level to prevent information leakage as we dealt with monozygotic twin pairs in our dataset, whose retinal images may hold strong resemblance. 
For training the filter, images of 26 2D modalities were used, excluding 3D OCT of the macula and ONH for the volumetric nature of the data and excluding widefield FSLO recordings for the small dataset size of 74 autofluorescence and 76 red-green images. Furthermore, images without known AmyloidPET status were included for training the filter. Some eyes had duplicate FAF images with good image quality. In this case, one image per eye per modality was selected for classifier development. Duplicate images could be used for filter development and generative model development.

\begin{figure}[H]
\centering
\includegraphics[width=\linewidth]{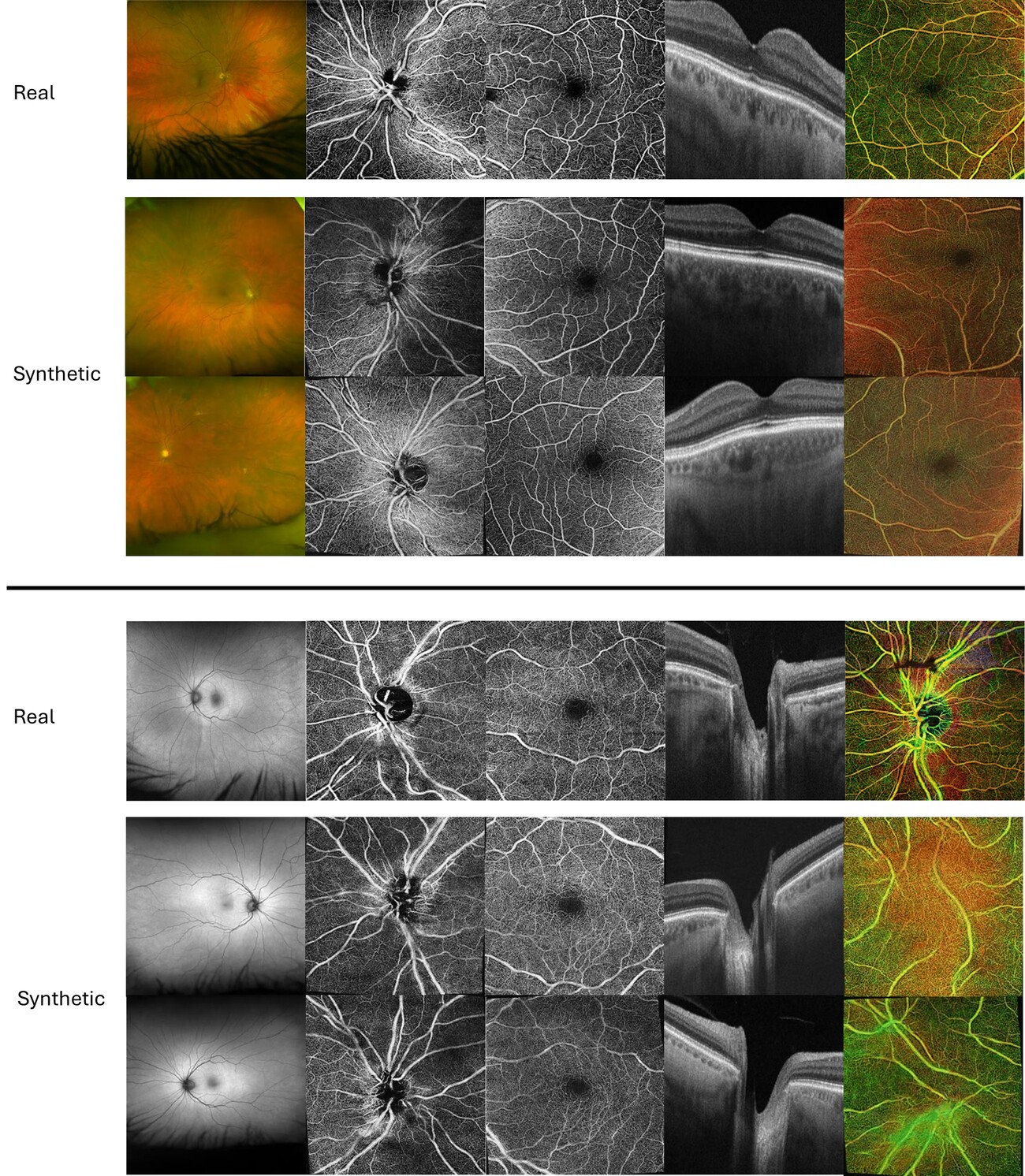}
\caption{Examples of generated synthetic images and real images for eight modalities. Top row from left to right: FRG, OCTA-SONH, OCTA-SMAC; OCT-BMAC; OCTA-EMAC. Bottom row left to right: FAF; OCTA-DONH; OCTA-DMAC; OCT-BONH; OCTA-EONH. The synthetic FAF images best resembled their real counterpart, with often accurate branching of the bloodvessels and even replication of the eye lashes at the periphery of the image. Synthetic FRG images failed to replicate the vasculature. Furthermore, the images were not sharp and most of them lacked accurate colors and would be green or yellow similar to the depth-encoded images. Most of the OCT-A images failed to replicate accurate branching of the blood vessels. OCT B-Scan images were overall quite realistic, although sometimes replication of one or two layers in the retina would occur. Depth-encoded OCT-A synthetic images were the least realistic, with malformations in the vasculature as well as in the coloring.}
\vspace{10em}
\label{fig:synthetic_images}
\end{figure}

\subsection*{Data extraction}
Extraction of the image data from the scanners involved decoding of binary files, anonymization of patient identifiers and association of files to the respective examinations. Anonymous patient identifiers were already created in the context of the two studies where the image data originate from. The eye examinations were stored in the scanners under these IDs. We replaced these identifiers with new pseudo-anonymous keys for the family, patient and eye such that our dataset cannot be directly related to information stored in the scanners.
Fundus SLO was extracted as DICOM files and directly anonymized. Image quality was reviewed based on movement or optical distortions such as lash or eyelid coverage and a straight eye gaze was required. Two examples of excluded FAF scans are shown in Figure \ref{fig:excluded_images}
Initially we intended to use widefield photos which allowed for a more complete depiction of the fundus, but this dataset contained only a small set of images with 74 autofluorescence and 76 red-green images. OCT-A images from the Zeiss Angioplex were evaluated by focus and resolution. Images were exported as .bmp and converted to jpg while sensitive metadata such as study date were removed from the filename. Exported files from the Heidelberg OCT lacked any reference to the examination it contained. Furthermore, these files were encoded as .E2E binary files, for which the scanner manufacturer provides no software to read the data. Therefore we had to decode the files ourselves. By adapting code from \url{https://github.com/marksgraham/OCT-Converter} to the structure of our data files we managed to extract binary image data, convert it into .npy files and extract patient identifiers with image metadata that allowed for identification of the recording type and follow-up order of examinations as multiple visits per patients were recorded.
\subsection*{Preprocessing}
After building the database, images were preprocessed to be used in the neural networks. Preprocessing was the same for the generative models, the filter and the classification models. Pixel values were normalized according to the untrained or pretrained neural networks (ResNet, EfficientNet). In case grayscale images were used on networks designed for RGB, grayscale channels were duplicated to create three-channel tensors. Age inputs were rescaled by 0.01 and sex was encoded as binary 0 (Male) 1 (Female). Fundus SLO images (originally $4000\times4000$ pixels) were cropped to remove the black background as visible in Figure \ref{fig:excluded_images} and then zoomed in to bring the fundus in the field of view and remove the eye lashes. Background and irrelevant structures of the sclera in B-scan OCT images (originally $1536\times1536$ pixels) were removed by cropping the top and bottom regions with pixel intensity lower than a manually set threshold. All 2-dimensional images were resized to $256\times256$ pixels. 3D OCT images were resized to $32\times128\times128$ (raw images: $73\times384\times496$ (OCT-VONH); $49\times512\times496$ (OCT-
VMAC)) 
\input{Figures/excluded_images/excluded_images}
\subsection*{Augmentation}
For augmenting the 2D images, random affine (shear, rotate) transformations, color (brightness, contrast), crop and zoom were applied. The zoom factor for B-Scan OCT was left unchanged as interpolation methods resulted in poor image quality. Volumetric 3D OCT was augmented by randomly cropping $75-10\%$ of the original volume in the X-Y plane. 

%% file: Figures/excluded_images/excluded_images.tex
\begin{figure}[H]
     \centering
     \begin{subfigure}[b]{0.49\textwidth}
         \centering
         \includegraphics[width=\textwidth]{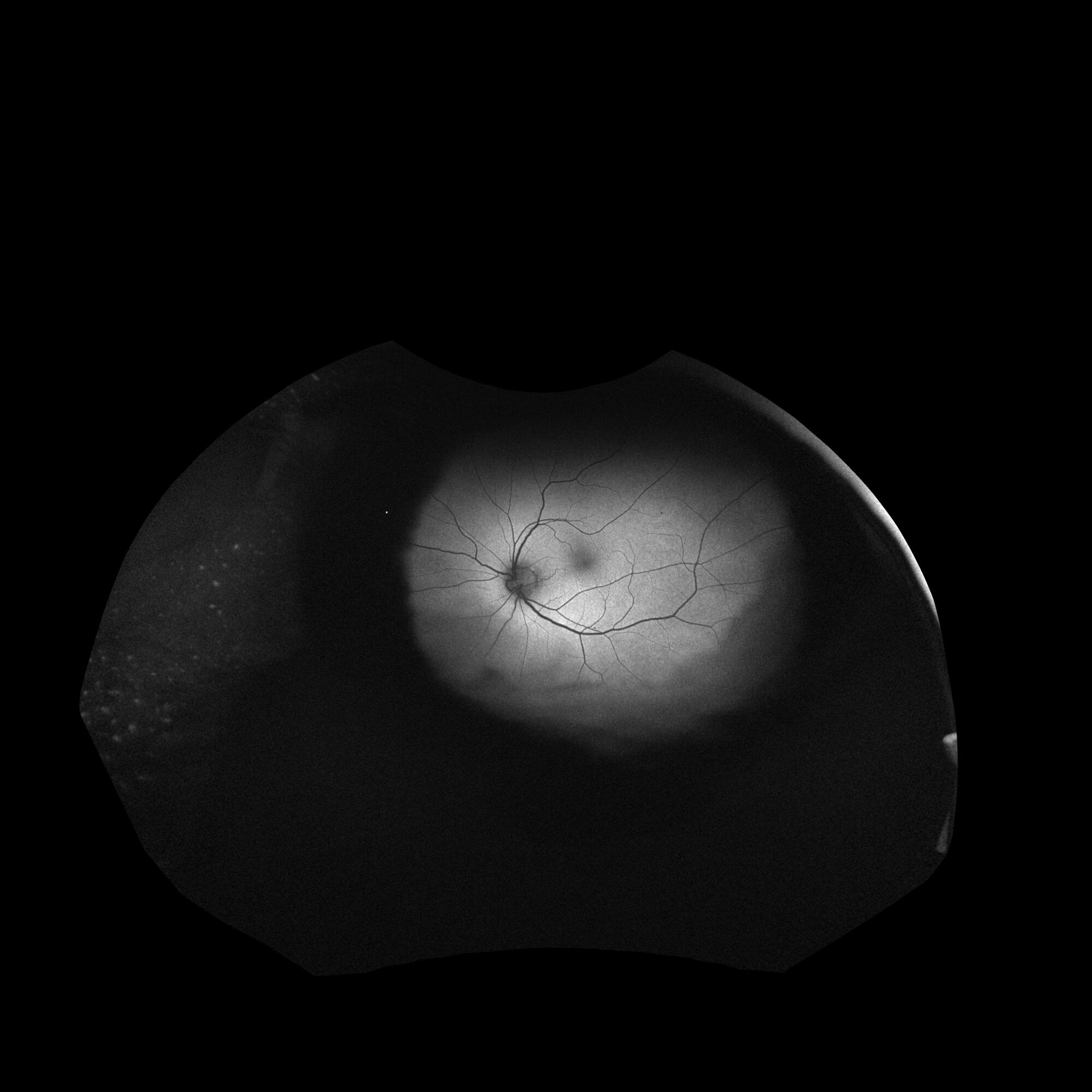}
         \caption{ }
     \end{subfigure}
      \begin{subfigure}[b]{0.49\textwidth}
         \centering
         \includegraphics[width=\textwidth]{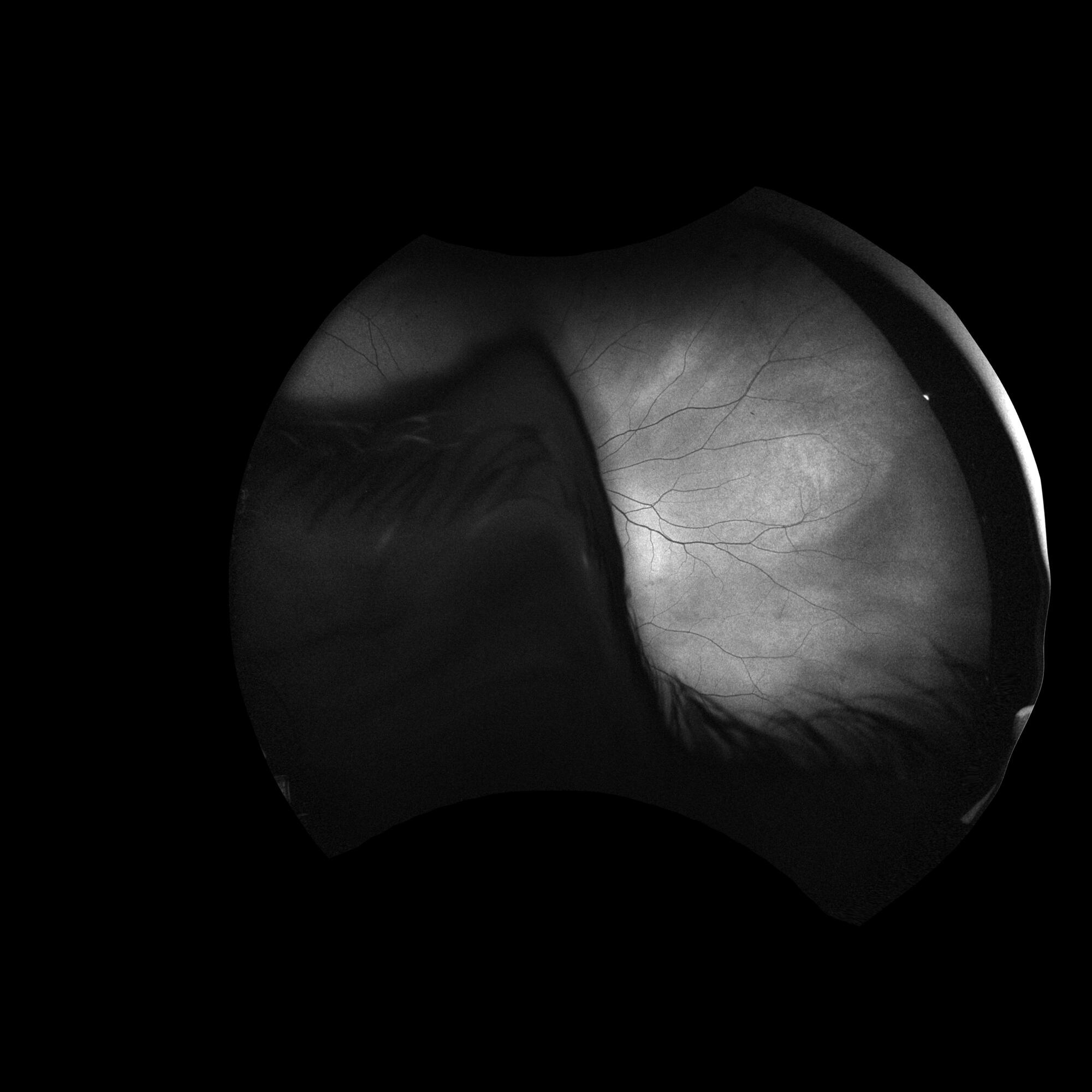}
         \caption{ }
     \end{subfigure}
    \caption{Example of excluded fundus. The two most often occurring reasons for exclusion were insufficient focus (a) and coverage of the fundus by the eyelid or eye lashes (b). }
    \label{fig:excluded_images}
\end{figure}

%% file: Supplementary/Models.tex
\subsection*{Neural networks}
The best learning rate, learning rate scheduler and optimizer hyperparameters were obtained through hyperparameter optimization. Model designs and approaches were also selected by comparison of small (25-trial) hyperparameter optimization experiments. Hyperparameter optimization used the Tree-structured Parzen Estimator algorithm through the Optuna Framework. Furthermore, training batches were formed with weighted random sampling to address class imbalance of AmyloidPET labels by oversampling the AmyloidPET+ samples.\newline \newline
All classification models were trained with early stopping: the training process was terminated when evaluation on the validation set showed no improvement. All AmyloidPET classifiers were trained with Focal Loss to emphasize on hard, misclassified examples.\parencite{FocalLoss} The models were trained to output the probability for AmyloidPET- status. \newline \newline
As the model was trained for outputting the probability for AmyloidPET-, these metrics were calculated with inverse ground truth labels and predictions. We calculate the predictions for AmyloidPET+ as $1 - outputs$ and the groundtruth labels for AmyloidPET as $1- groundtruth$, this way the sensitivity and specificity reflected the model's performance with respect to detecting AmyloidPET+ cases.
\newline \newline
The classification models were evaluated with metrics derived from receiver-operator (ROC) and precision-recall (PR) curves. Precision and recall were determined based on a threshold obtained with the youden's index.

\subsubsection*{DDPM}
%

 
GANs consist of two competing models, a generator produces fake images and a discriminator distinguishes between real and fake images. In this process the generator is secluded from the real images, so there is little risk of the generator producing copies of real images. unlike GANs, which prioritize image fidelity over diversity and seclude the training images from the generator, DDPMs are prone to memorization as generator has access to information of the training images. This is because the loss is computed between the predicted noise and noise to be removed in the real image. DDPMs initially map an input image to a noise image by gradually adding Gaussian noise in many small steps. Subsequently, the reverse process is learned in small steps by predicting what noise should be removed from a given image at a certain step in the denoising process. A popular implementation of DDPMs involves a Unet-shaped CNN for predicting the noise to be removed. The selected model to produce images corresponding tot specific AmyloidPET status was a conditional U-Net DDPM. This model performs conditioning on the timestep embedding that is used for predicting the noise that is to be removed. This conditioning is achieved through addition of the class label embedding vector to the time step embedding vector. 
\newline \newline 
Different configurations for attention levels, filter channels, residual connections and architecture variations such as latent diffusion networks and ControlNets were tested. A conditional DDPM with U-Net backbone was used from the MONAI Generative open-source project.\parencite{MONAI} The model was trained to take inputs of $256 \times 256$ pixels with three blocks in the encoders and decoders $(64, 128$ and $128$ channels$)$, two residual blocks per encoder/decoder block and spatial transformer attention mechanisms with 32 channels per head in the last encoder/decoder block. The spatial transformer learns to model complex spatial transformations to align the feature maps more accurately with the target distribution. The diffusion model was trained to produce images corresponding to the desired AmyloidPET status by conditioning the U-Net and thereby conditioning the denoising process during training and sampling. This conditioning was achieved through incorporation of the class label embedding into the timestep embedding. The timestep embedding is incorporated into the output by passing it through a linear layer and subsequent summation with the intermediate representation of the input at each block of the UNet. We used a scaled linear beta noise schedule function to add noise to the training images, initial learning rate of $1e^{-3}$, cosine annealing learning rate scheduler, and Adam optimizer without weight decay. 
\newline

\subsubsection*{Filter}
The network for the filter was an EfficientNetB0 model pretrained on ImageNet. All layers in the model were finetuned during training. The network was trained for 100 epochs with learning rate $1e^{-5}$, Adam optimizer with weight decay of 0.1 and no learning rate scheduler. The training objective was to correctly recognize the modality of an image out of 26 different classes. The model performed well, with Matthew's Correlation Coefficient (MCC) of 0.9900 for the predictions on the validation set and MCC of 0.9968 on the test set. Results for all modalities are shown in Table \ref{tab:filter_accuracy} and Figure \ref{fig:filter_confusion_matrix}. We tried applying a filter with lower validation accuracy, by training for fewer epochs, in the hope that an image would have to represent its modality very strongly and therefore unrealistic images would be discarded more often. However, this did not make any difference. 

\input{Tables/filter_evaluation_t} 

\input{Figures/filter_confusion_matrix_t}

\subsubsection*{Unimodal classifiers}
ResNet and the EfficientNet family architectures were explored in combination with and without pre-trained ImageNet weights as starting points. In addition to modality-specific classifiers we also experimented with training one shared modality-aware classifier with feature-wise linear modulation (FiLM) but this did not perform well.\parencite{FiLM} \newline \newline
The starting point for the unimodal classifiers was a EfficientNetB0 network without pretrained weights. The final layer was changed to a binary prediction layer. Classifiers were trained for 200 epochs with various initial learning rates, a step LR scheduler and Adam optimizer without weight decay. Binary cross entropy was compared with Focal Loss in 20-trial hyperparameter experiments. As the loss stopped converging throughout the 200 epochs we selected model checkpoints taken from the best epoch.
\newline 
\subsubsection*{Multimodal classifiers}
The multimodal classifier is a fusion of the unimodal classifiers through fully connected layers for combining unimodal predictions and metadata. We experimented with various fusion methods, none of these fusion methods outperformed our selected approach (i.e. heterogeneous fusion). The methods we experimented with were:
\begin{itemize}
 \item Fusion of embedding vectors produced by the unimodal classifiers by transforming the vectors into 2-dimensional arrays and applying convolution operations on the resulting multi-channel vectors and several fully connected layers at the end.
 \item Fusion of embedding vectors produced by the unimodal classifiers through multiple fully connected layers. 
 \item Early or late fusion of metadata with classification features through either incorporation of metadata early in the fully connected network or late in the fully connected network.
\end{itemize}
As the loss on the validation set stopped converging but would not deteriorate throughout the 200 epochs, we would select the model checkpoint from the last epoch.
\newline
\subsubsection*{Modality-aware classifier with FiLM}
As the filter model successfully distinguished modalities, we experimented with training one modality-aware classification network that was capable of predicting AmyloidPET status for any of the four modalities through an additional input that informed the model of the modality type. This additional input was an embedding vector extracted from the first fully connected layer of the filter model. Figure \ref{fig:TSNE_figures} displays these embeddings on reduced 2-dimensional axes for all images in the test set and demonstrates that this embedding was suitable for distinguishing modality types. Although FiLM-based classification performed poorly, we show the filter embeddings in Figure \ref{fig:TSNE_figures} to demonstrate that AmyloidPET status was not correlated in the embedding space and therefor the filter was not biased towards a specific AmyloidPET group. \newline \newline The embeddings were used in the modality-aware classifier by conditioning intermediate activations of the classification networks in FiLM layers. Through FiLM, the intermediate outputs of selected layers of a neural network are transformed with scale and bias parameters. These parameters are mapped from a given conditioning embedding, in our case the filter embedding. The mapping from embedding vector to a scale and bias parameter is learned during model training, as part of the same back-propagation chain as the rest of the model training. As a result, the scale and bias parameters learn to modulate the model operations according to the embedding vector that describes the image modality. We added two FiLM layers in each network: one between the stem and Block 1 and between Block 2 and Block 3. 
 
\input{Figures/TSNE_figures_t}

%% file: Tables/filter_evaluation_t.tex
\begin{table}[H]
\centering
\begin{tabular}{lllll}
Modality (abbreviation) & Modality (explanation)                     & precision & recall & f1-score \\
COL                     & Fundus SLO red-green                   & 1.00      & 1.00   & 1.00     \\
FAF                     & Fundus SLO autofluorescence            & 1.00      & 1.00   & 1.00     \\
OCTA-EMAC               & OCT-A Macula Angiography Depth Encoded & 1.00      & 1.00   & 1.00     \\
OCTA-EONH               & OCT-A ONH Angiography Depth Encoded    & 1.00      & 1.00   & 1.00     \\
OCTA-WONH               & OCT-A ONH Angiography WholeEye         & 0.99      & 1.00   & 0.99     \\
OCTA-WMAC               & OCT-A Macula Angiography WholeEye      & 1.00      & 1.00   & 1.00     \\
OCTA-ORCCMAC            & OCT-A Macula Angiography ORCC          & 1.00      & 1.00   & 1.00     \\
OCTA-ORCCONH            & OCT-A ONH Angiography ORCC             & 1.00      & 1.00   & 1.00     \\
OCTA-RMAC               & OCT-A Macula Angiography Retina        & 0.95      & 1.00   & 0.98     \\
OCTA-RONH               & OCT-A ONH Angiography Retina           & 1.00      & 1.00   & 1.00     \\
OCTA-DMAC               & OCT-A Macula Angiography Deep          & 1.00      & 0.99   & 0.99     \\
OCTA-DONH               & OCT-A ONH Angiography Deep             & 1.00      & 1.00   & 1.00     \\
OCTA-SMAC               & OCT-A MaculaAngiography Superficial    & 1.00      & 0.96   & 0.98     \\
OCTA-SONH               & OCT-A ONH Angiography Superficial      & 1.00      & 1.00   & 1.00     \\
OCT-WONH                & OCT ONH Structure WholeEye             & 1.00      & 1.00   & 1.00     \\
OCT-WMAC                & OCT Macula Structure WholeEye          & 1.00      & 1.00   & 1.00     \\
OCT-ORCCMAC             & OCT Macula Structure ORCC              & 1.00      & 1.00   & 1.00     \\
OCT-ORCCONH             & OCT ONH Structure ORCC                 & 1.00      & 1.00   & 1.00     \\
OCT-RMAC                & OCT Macula Structure Retina            & 0.99      & 1.00   & 0.99     \\
OCT-RONH                & OCT ONH Structure Retina               & 1.00      & 1.00   & 1.00     \\
OCT-DMAC                & OCT Macula Structure Deep              & 1.00      & 0.97   & 0.99     \\
OCT-DONH                & OCT ONH Structure Deep                 & 1.00      & 1.00   & 1.00     \\
OCT-SMAC                & OCT Macula Structure Superficial       & 1.00      & 1.00   & 1.00     \\
OCT-SONH                & OCT ONH Structure Superficial          & 1.00      & 1.00   & 1.00     \\
OCT-BMAC                & OCT Macula B-Scan                      & 1.00      & 1.00   & 1.00     \\
OCT-BONH                & OCT ONH B-Scan                         & 1.00      & 1.00   & 1.00     \\
weighted avg            &                                        & 1.00      & 1.00   & 1.00   
\end{tabular}
\caption{Accuracy of filter}
\label{tab:filter_accuracy}
\end{table}

%% file: Figures/filter_confusion_matrix_t.tex
\begin{figure}[H]
\centering
\includegraphics[width=\textwidth]{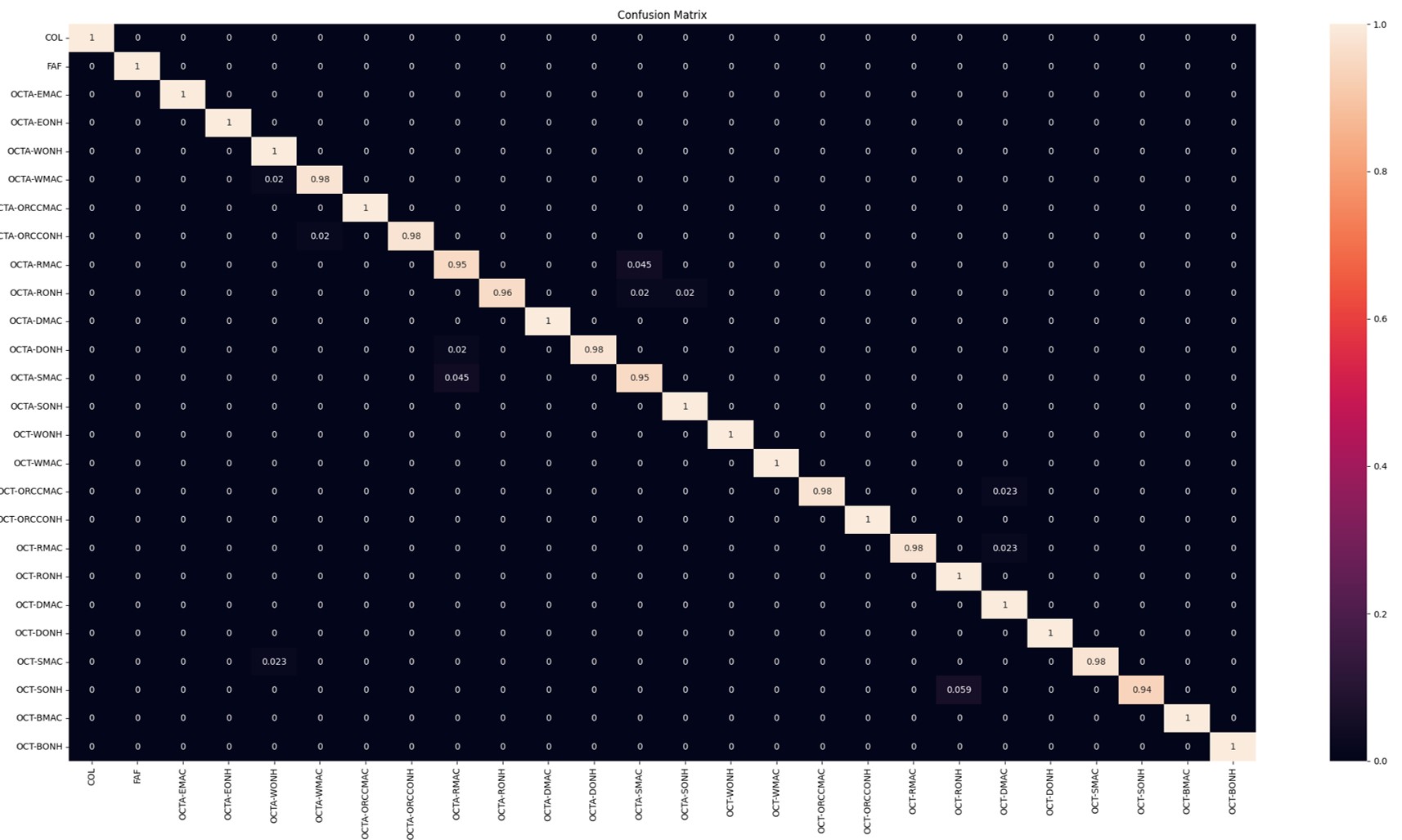}
\caption{Confusion matrix for distinguishing between 26 modalities by the filter. Wrong predictions happened among modalities that were relatively similar. }
\label{fig:filter_confusion_matrix}
\end{figure}

%% file: Figures/TSNE_figures_t.tex
\begin{figure}[H]
     \centering
     \begin{subfigure}[b]{0.8\textwidth}
         \centering
         \includegraphics[width=\textwidth]{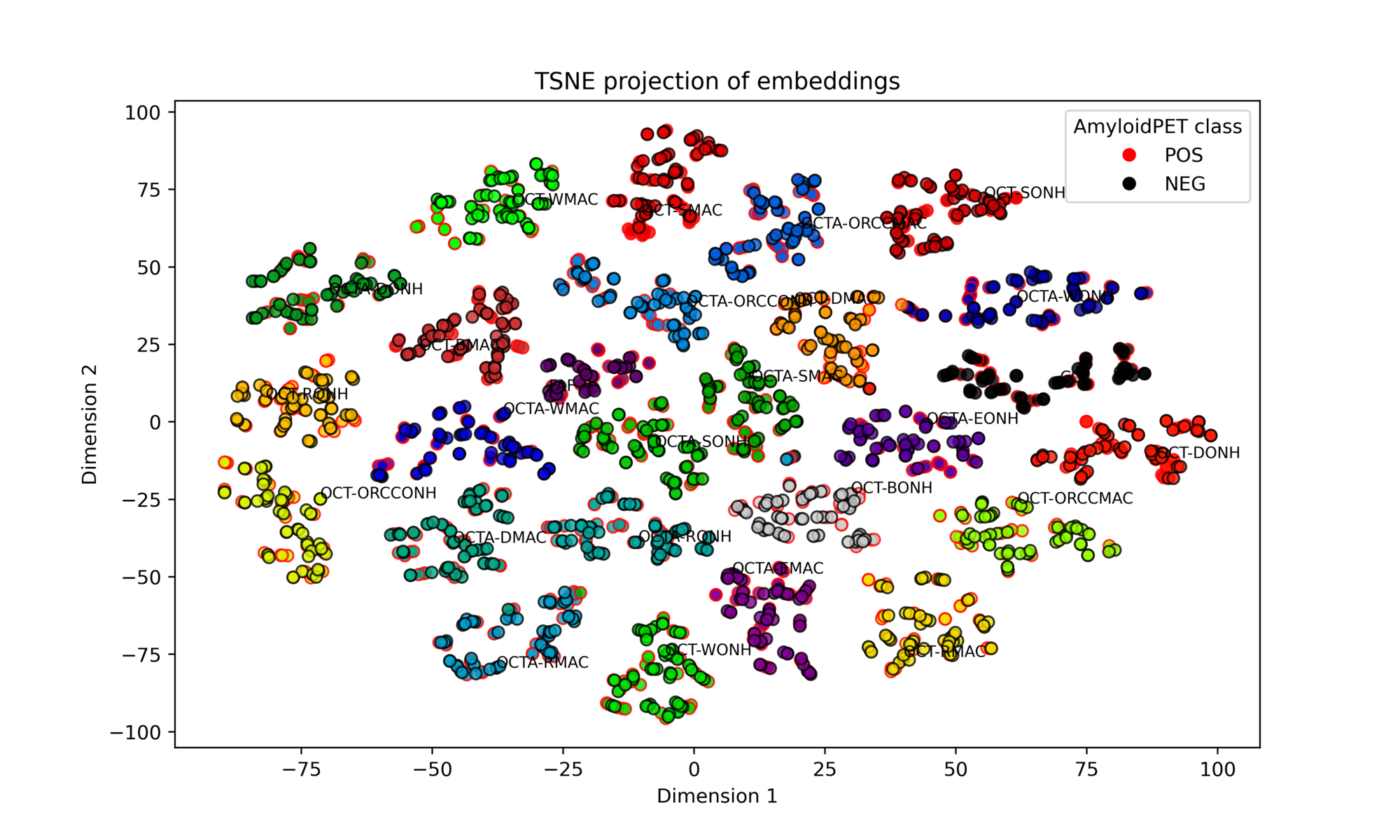}
         \caption{ }
     \end{subfigure}
     \begin{subfigure}[b]{0.8\textwidth}
         \centering
         \includegraphics[width=\textwidth]{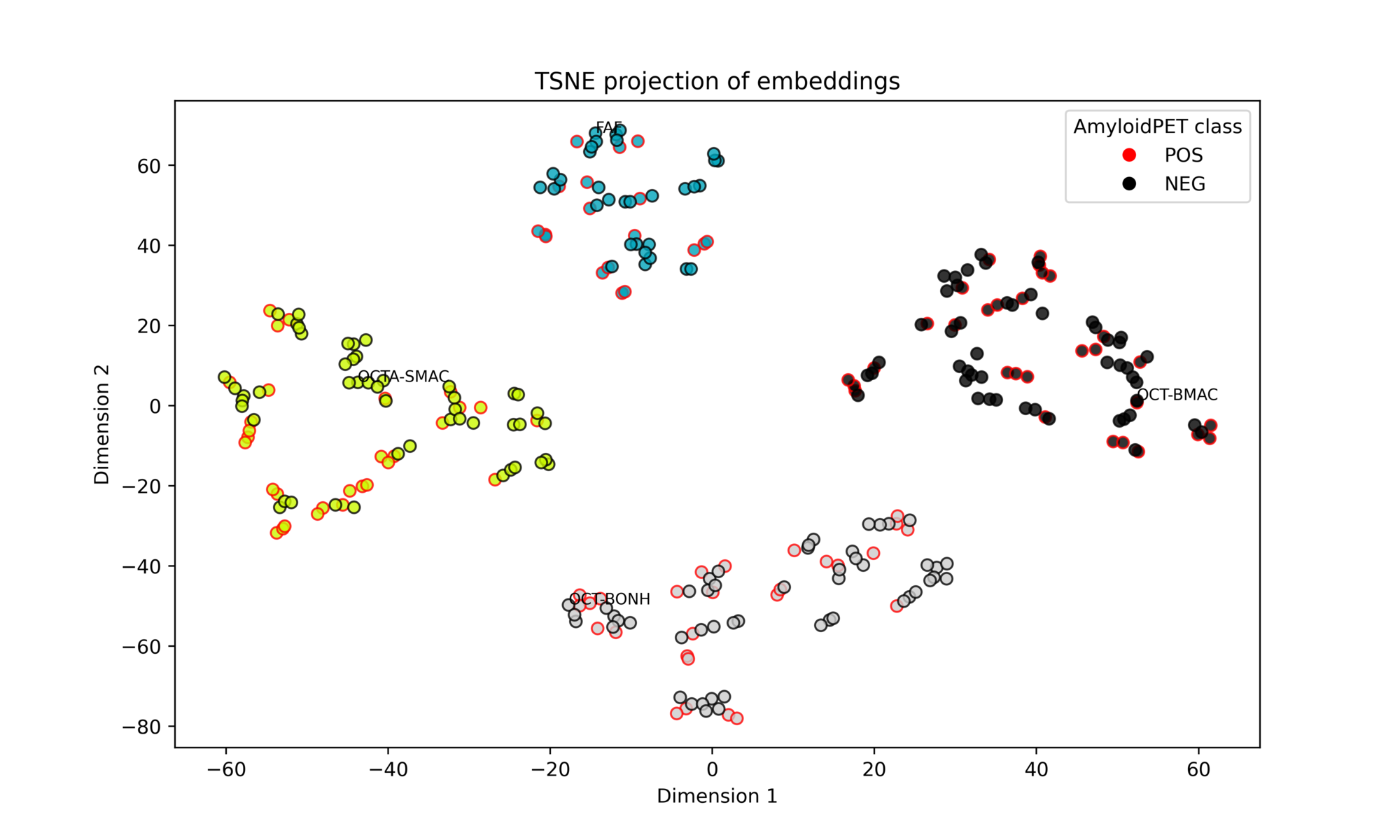}
         \caption{ }
     \end{subfigure}
    \caption{TSNE projection of filter embedding of (a) images of all modalities and (b) the modalities used in the classification experiments. Modalities can be distinguished by filling color. AmyloidPET status is distinguished by edge color, with which we want to demonstrate that the embedding space is not correlated to AmyloidPET status. TSNE embeds the points from a higher dimension to a lower dimension trying to preserve the neighborhood of that point by minimizing the Kullback–Leibler divergence between distributions with respect to the locations of the points in the map. Projections (a) and (b) come from TSNE calculations with different sets of modalities which results in different projections of the OCT-SMAC, OCT-BONH, OCT-BMAC and FAF points on the 2D spaces.}
    \label{fig:TSNE_figures}
\end{figure}

%% file: Supplementary/Cams.tex
\subsection*{Class activation maps}
We implemented Gradient-weighted CAM (GradCAM) with the library \texttt{pytorch-grad-cam} to produce heatmaps that give insight into the model's activations after the last convolutional layer.\parencite{GradCAM} Since the activity of convolutional layers often maps spatially to the input, we upsampled the GradCAM attributions to mask the input. GradCAM applied to our binary classification network computed the gradient of the binary output layer with respect to each of the network's activations at the selected convolutional layer. To produce the heatmap, the gradient at the output layer was computed and subsequently multiplied with the layer's activations. We depict GradCAMs as a heatmap in which the red-orange color regions were the most discriminative areas for the model to predict AmyloidPET- and the green-blue areas were the least salient. Figures \ref{fig:cams_real_classifiers}-\ref{fig:cams_synthetic_classifiers} contain the same retinal images, allowing for direct comparison of the heatmaps of different classifiers.
\newline \newline
The GradCAM heatmaps for the three different classifiers trained on OCTA-SMAC showed different patterns. The classifier trained on real images showed highest response to blood vessels in the periphery and there is a larger area of high response to real images than synthetic images. In contrast, the pretrained classifier showed a larger response area in the center of the image. The response to the synthetic images was different, with high responses to a large central area. The model trained on synthetic images displayed attention to small areas in the periphery. We did not observe large differences in the size of the heatmap areas with high response between true negative and true positive images, except for the heatmaps on real images produced by the classifier trained on synthetic images. This was also reflected by the small differences in output values. Attention of the OCT-BMAC classifiers did not show that clinically related features were learned by the model. The baseline FAF classifier attended to the periphery of the fundus. \newline \newline
The OCT-BONH classifier trained on real images showed high response to small areas in the layers of the retina. The pretrained classifier showed responses to similar areas, with slightly larger areas of high activation in the pretrained images. The shape and location of the high response areas implied that these models learned to identify meaningful features, however the output values for the different classes were very close to each other ($0.494$ and $0.485$ compared to $0.471$ and $0.489$) which implies that the classes were not well distinguished by these features. The classifier trained on synthetic data showed larger areas of high response, even more so in the real images compared to synthetic images. In contrast, the pretrained FAF classifier identified areas around the fovea and ONH. 
\newline \newline
The pretrained OCT-BMAC classifier showed very little response in any of the images, except for the heatmap on the real AmyloidPET+ images which depicted a localized response in the photoreceptor layer and retinal pigment epithelium. Interestingly, the classifier trained on real images showed localized response to a synthesis artifact in the synthetic AmyloidPET- images. The heatmaps produced by this classifier seemed different for the images, nonetheless the confidence for AmyloiPET- was similar. In contrast, the classifier trained on synthetic images showed a low response to the synthesis artifact. The responses were more localized than in the classifier trained on real images.
\newline \newline
The GradCAM heatmaps for the three different classifiers trained on FAF showed different patterns. The pretrained model identified areas around the fovea and ONH. For both synthetic and real images there were slightly larger areas of high response which was also reflected in the output scores. The model trained on synthetic images showed responses of very different shapes when comparing the synthetic and real images. The synthetic images showed a large squared area of high response whereas the response to real images was more restricted to specific areas, mostly in the periphery of the fundus. The model trained on real images showed strongest responses in small areas of the far periphery of the fundus. This model produced larger areas of high response to the synthetic images. These responses were more localized around the ONH and fovea.

\input{Figures/cams/cams_real}
\input{Figures/cams/cams_finetuned}
\input{Figures/cams/cams_synthetic}

%% file: Figures/cams/cams_real.tex
\begin{figure}[H]
     \centering
     \begin{subfigure}[b]{0.48\textwidth}
         \centering
         \includegraphics[width=\textwidth]{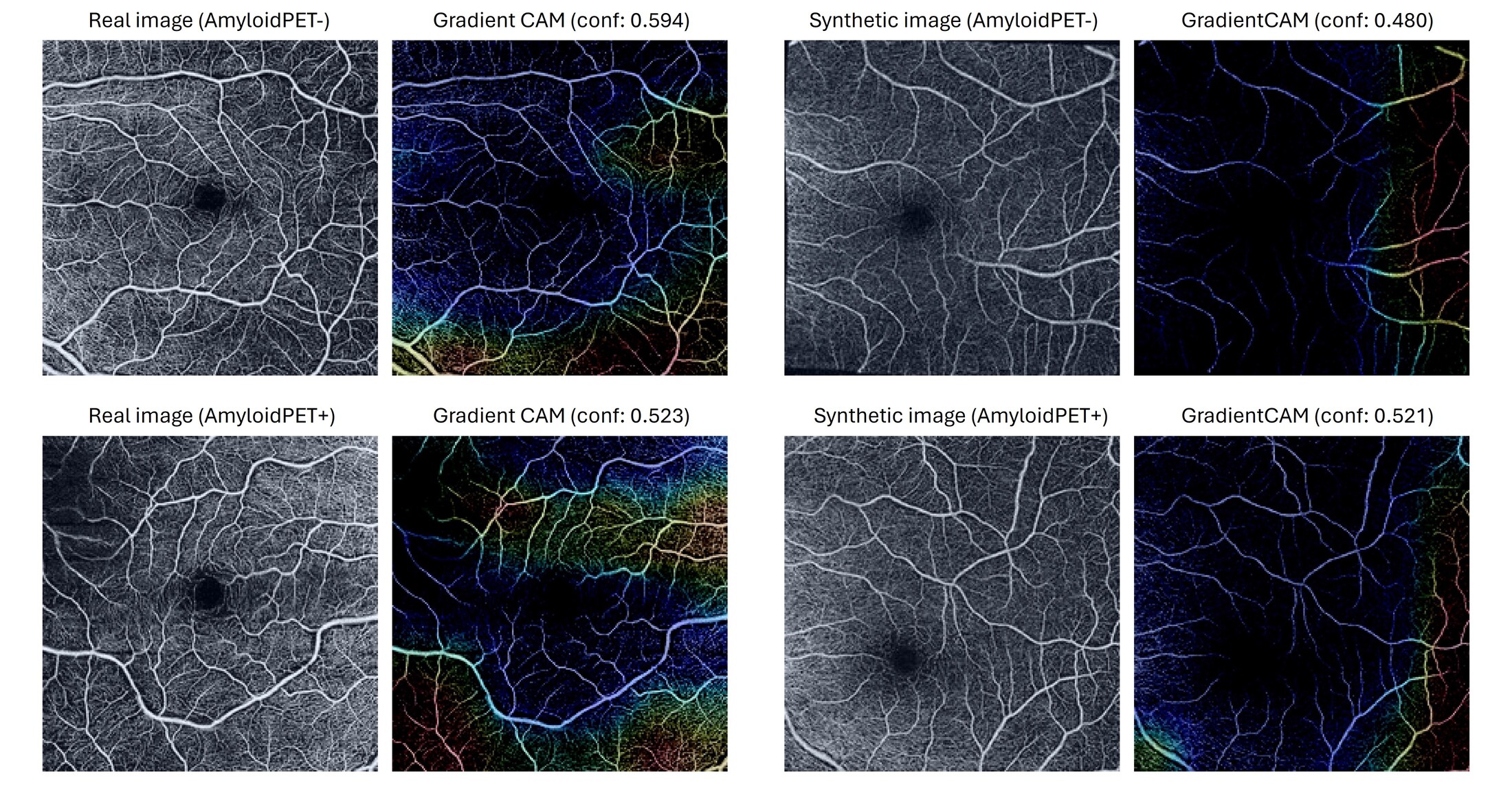}
         \caption{ }
     \end{subfigure}
      \begin{subfigure}[b]{0.48\textwidth}
         \centering
         \includegraphics[width=\textwidth]{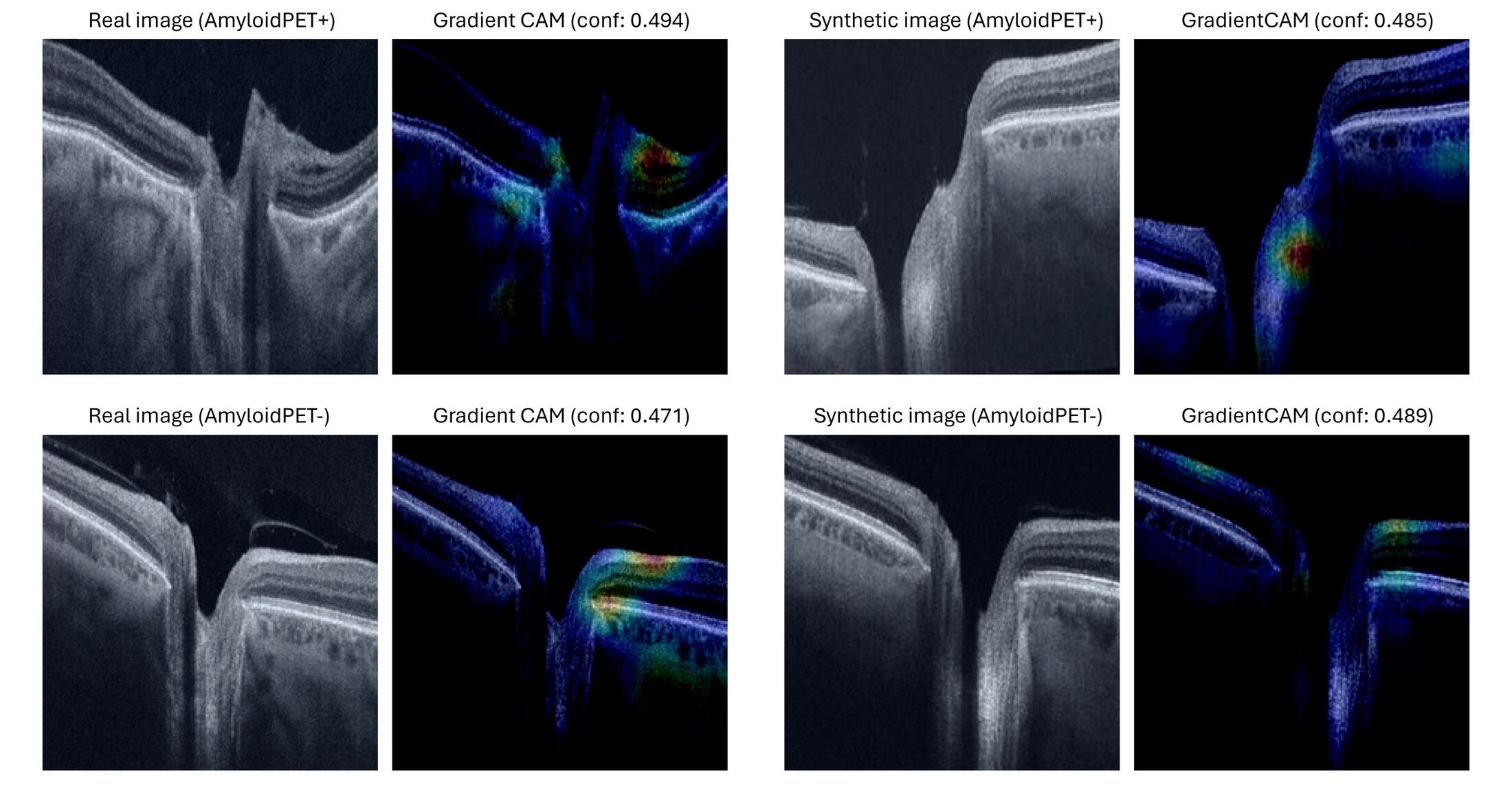}
         \caption{ }
     \end{subfigure}
     \begin{subfigure}[b]{0.48\textwidth}
         \centering
         \includegraphics[width=\textwidth]{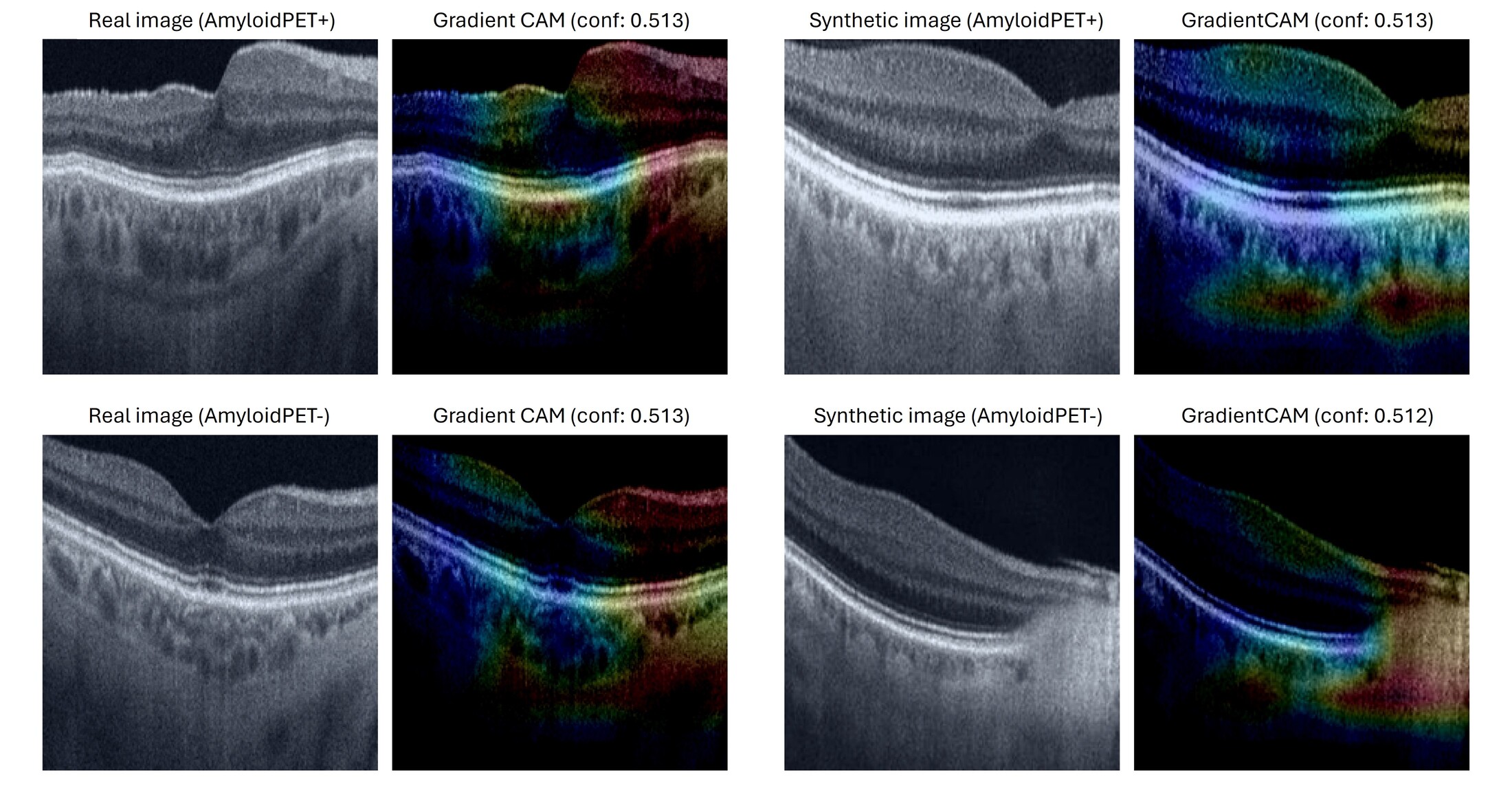}
         \caption{ }
     \end{subfigure}
     \begin{subfigure}[b]{0.48\textwidth}
         \centering
         \includegraphics[width=\textwidth]{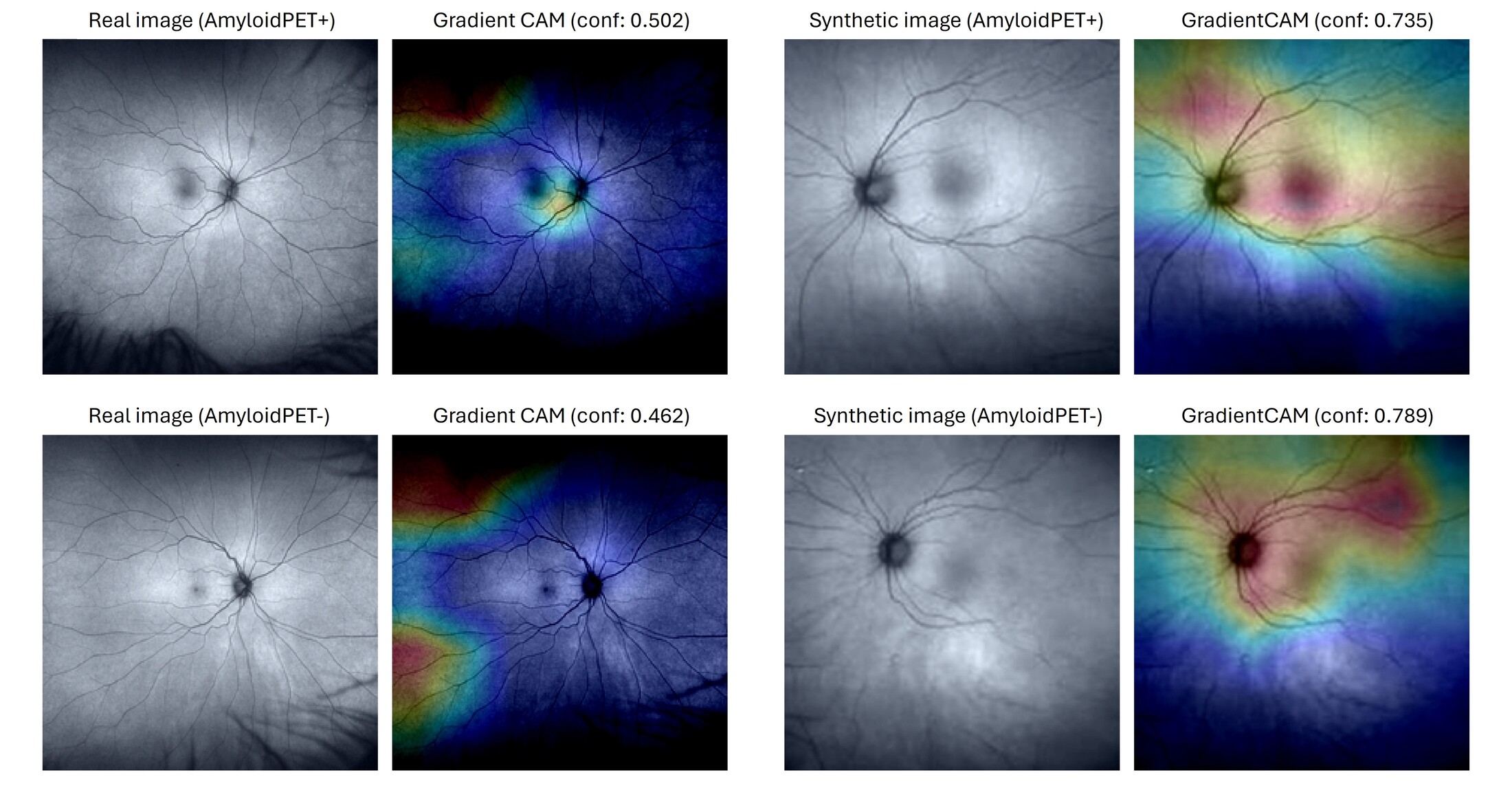}
         \caption{ }
     \end{subfigure}
    \caption{GradCAMs generated for classifiers trained on real images only. The subfigures show pairs of input images and GradCAM heatmaps for OCTA-SMAC (a), OCT-BMONH (b), OCT-BMAC (c), FAF (d). Each subfigure has four images. Left top to bottom: real images for AmyloidPET+ and AmyloidPET-. Right top to bottom: synthetic Images for AmyloidPET+ and AmyloidPET-.}
    \label{fig:cams_real_classifiers}
\end{figure}

%% file: Figures/cams/cams_finetuned.tex
\begin{figure}[H]
     \centering
     \begin{subfigure}[b]{0.48\textwidth}
         \centering
         \includegraphics[width=\textwidth]{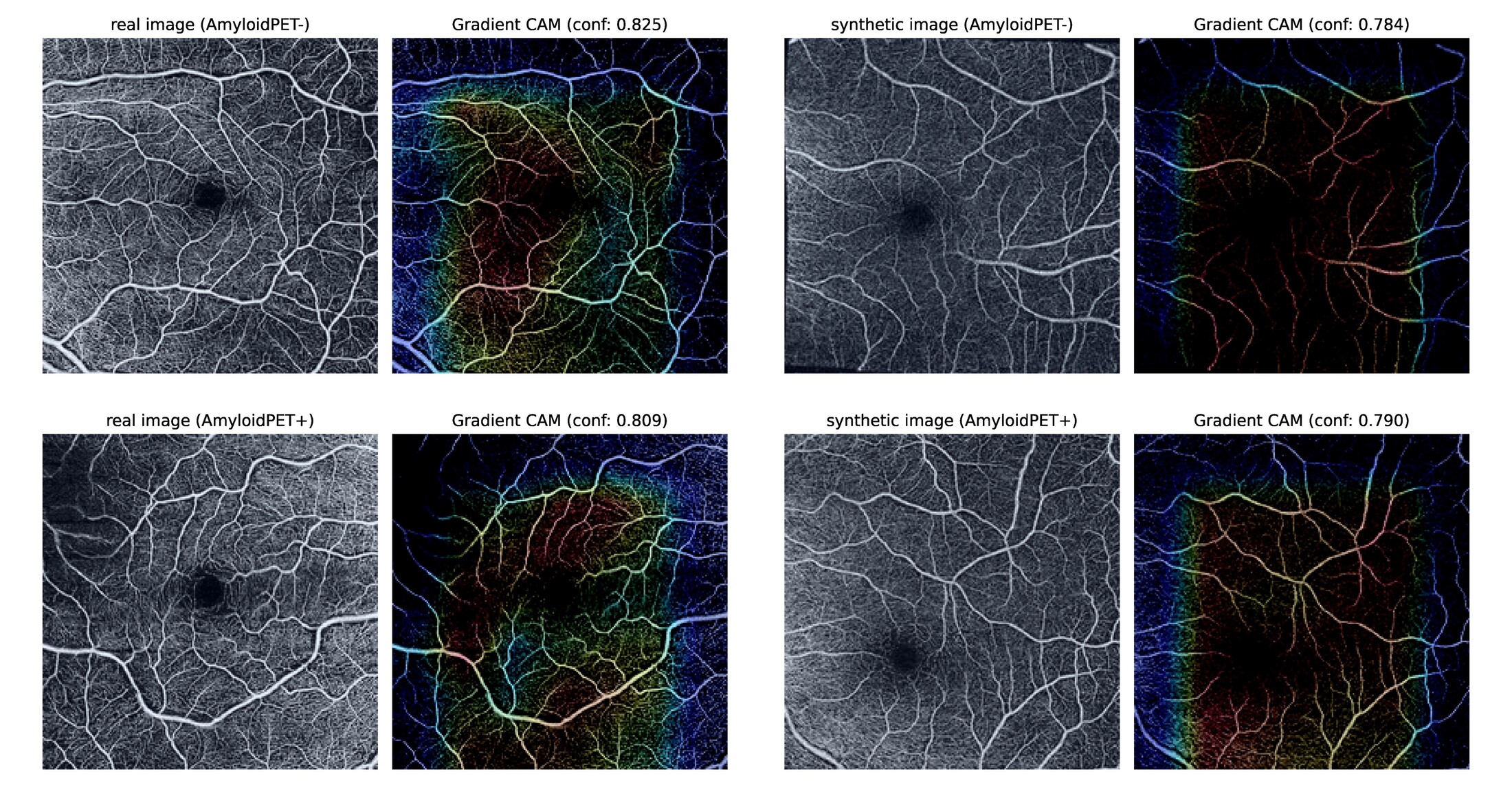}
         \caption{ }
     \end{subfigure}
      \begin{subfigure}[b]{0.48\textwidth}
         \centering
         \includegraphics[width=\textwidth]{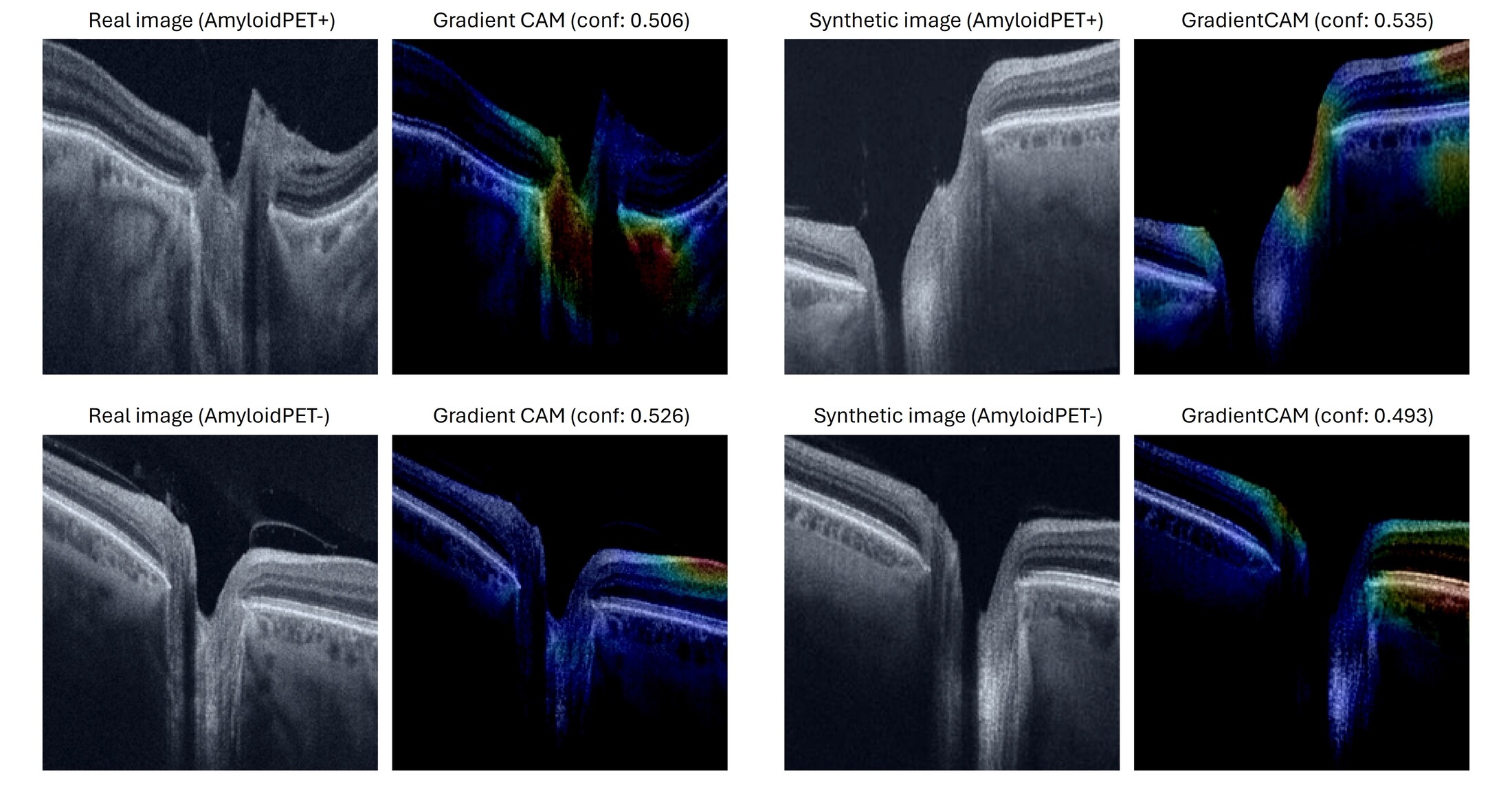}
         \caption{ }
     \end{subfigure}
     \begin{subfigure}[b]{0.48\textwidth}
         \centering
         \includegraphics[width=\textwidth]{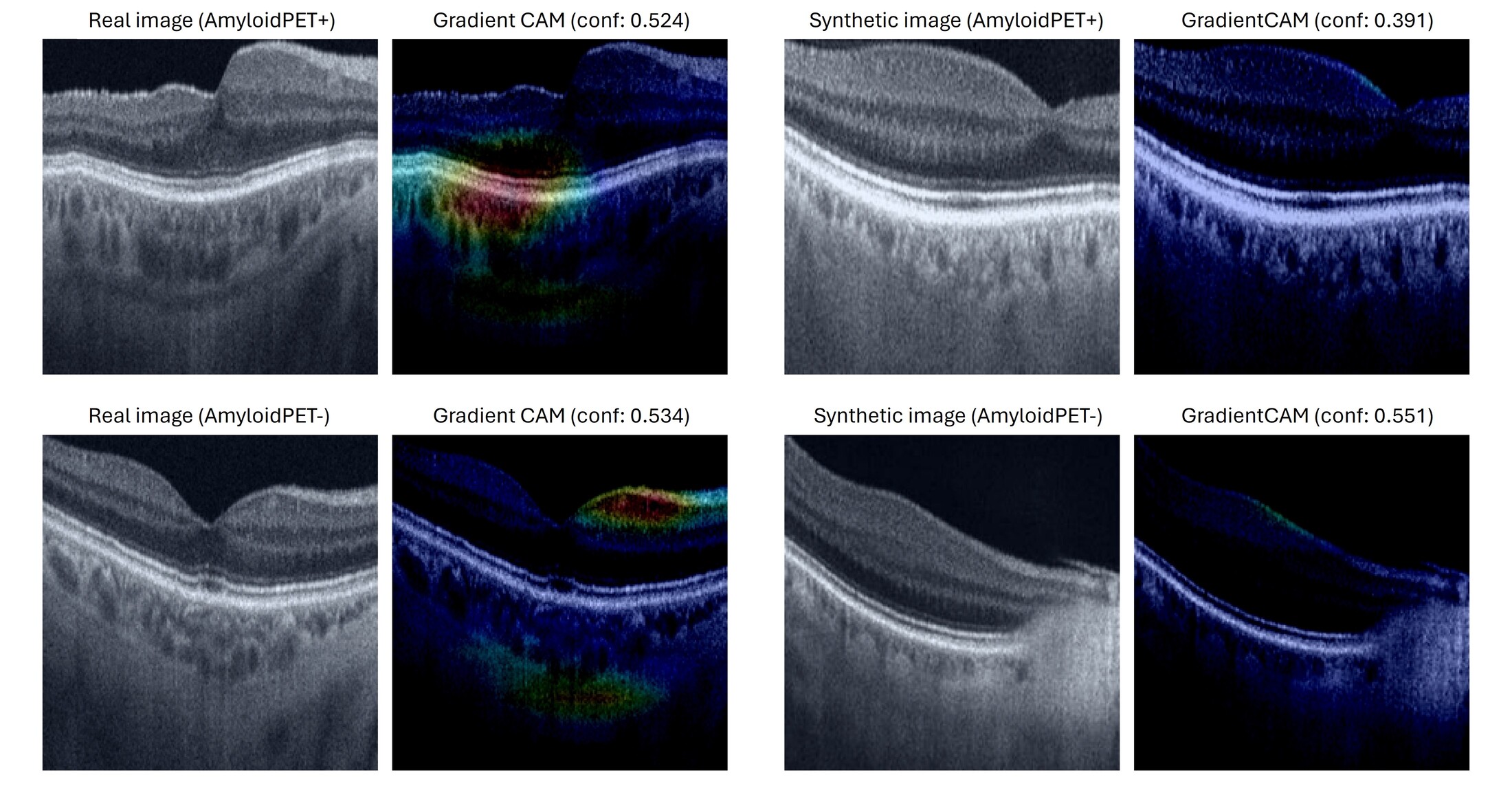}
         \caption{ }
     \end{subfigure}
     \begin{subfigure}[b]{0.48\textwidth}
         \centering
         \includegraphics[width=\textwidth]{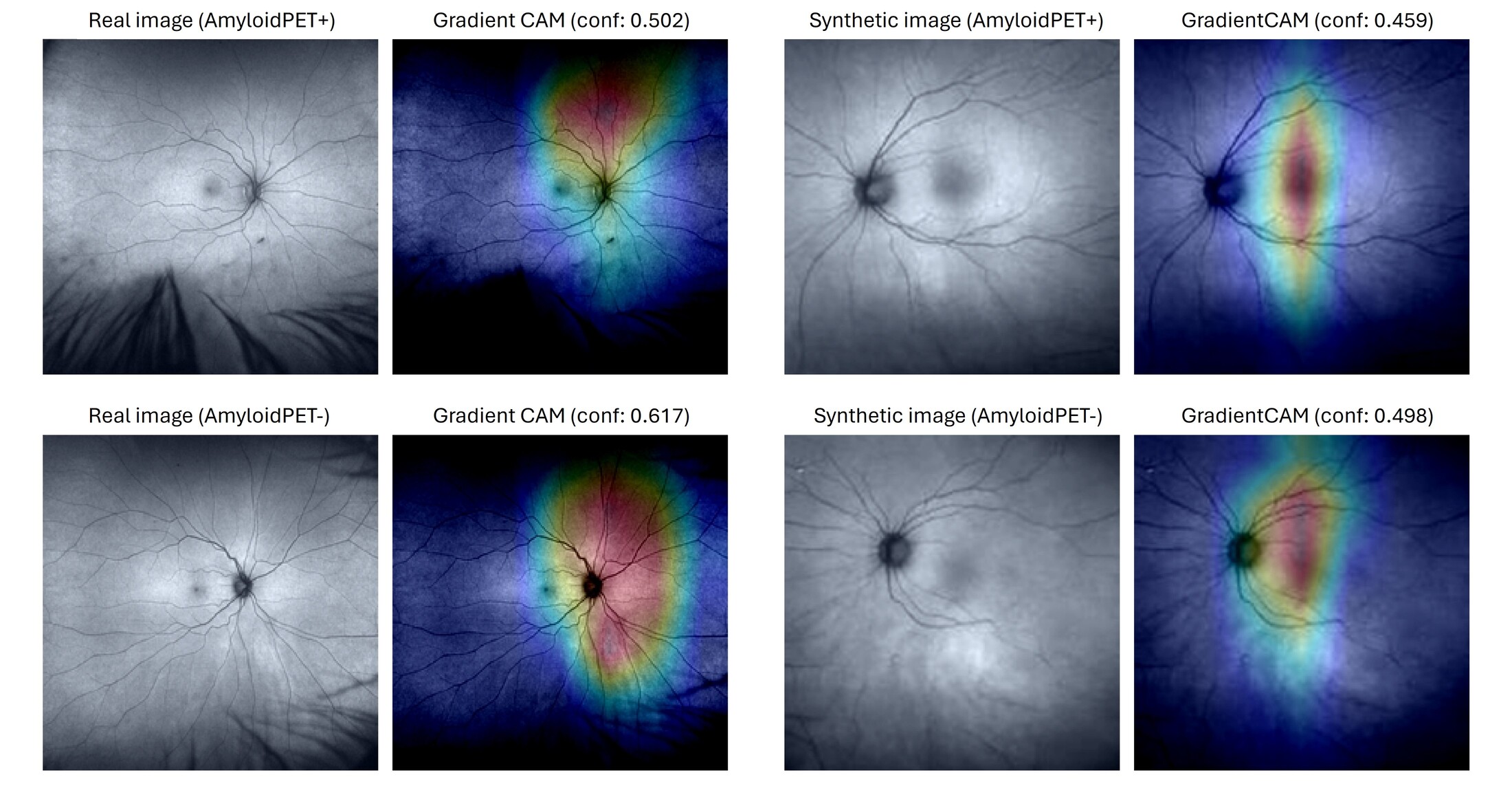}
         \caption{ }
     \end{subfigure}
    \caption{GradCAMs generated for classifiers pretrained on synthetic images and finetuned on real images. The subfigures show pairs of input images and GradCAM heatmaps for OCTA-SMAC (a), OCT-BMONH (b), OCT-BMAC (c), FAF (d). Each subfigure has four images. Left top to bottom: real images for AmyloidPET+ and AmyloidPET-. Right top to bottom: synthetic Images for AmyloidPET+ and AmyloidPET-.  }
    \label{fig:cams_pretrained_classifiers}
\end{figure}

%% file: Figures/cams/cams_synthetic.tex
\begin{figure}[H]
     \centering
     \begin{subfigure}[b]{0.48\textwidth}
         \centering
         \includegraphics[width=\textwidth]{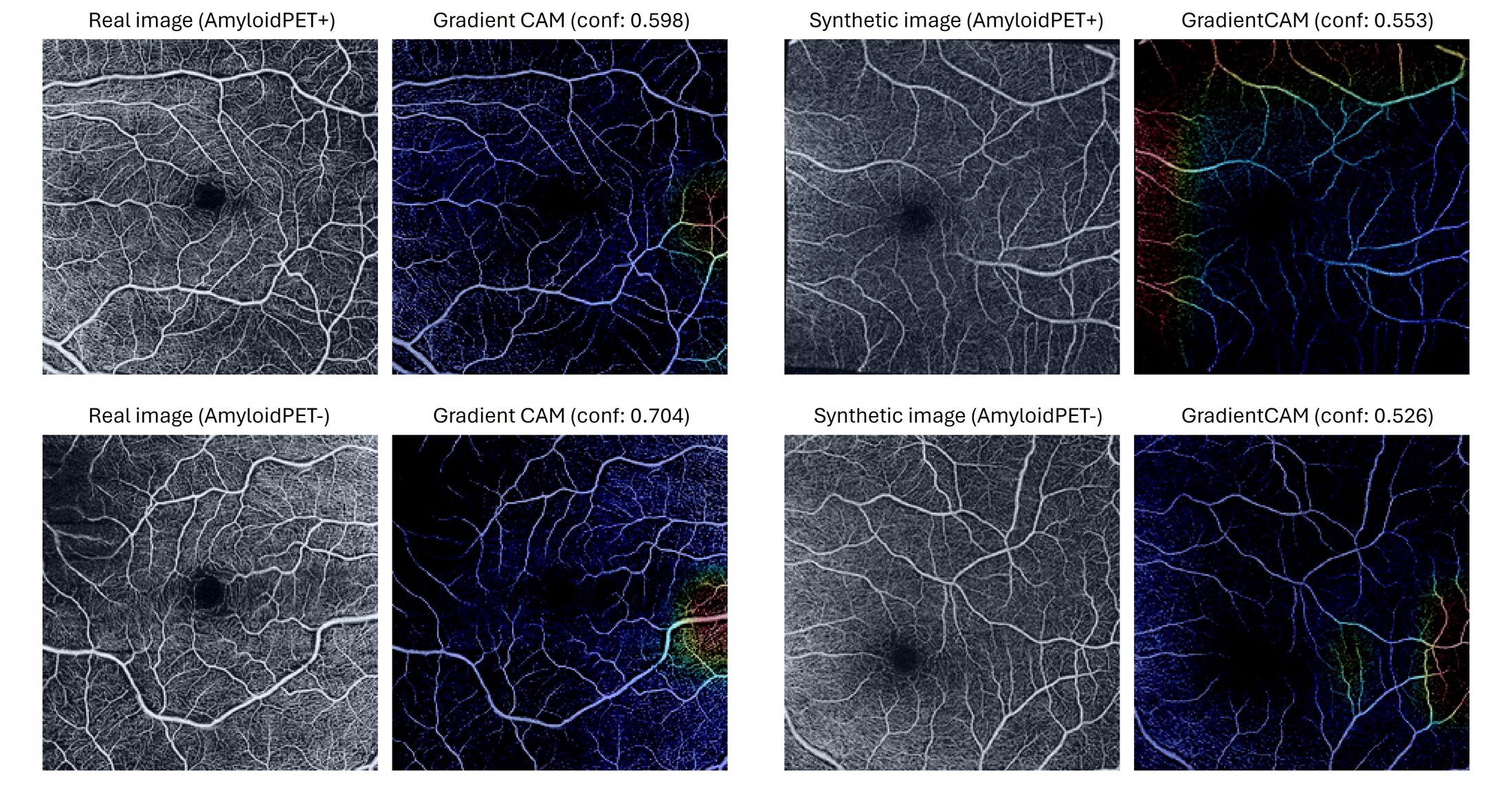}
         \caption{ }
     \end{subfigure}
      \begin{subfigure}[b]{0.48\textwidth}
         \centering
         \includegraphics[width=\textwidth]{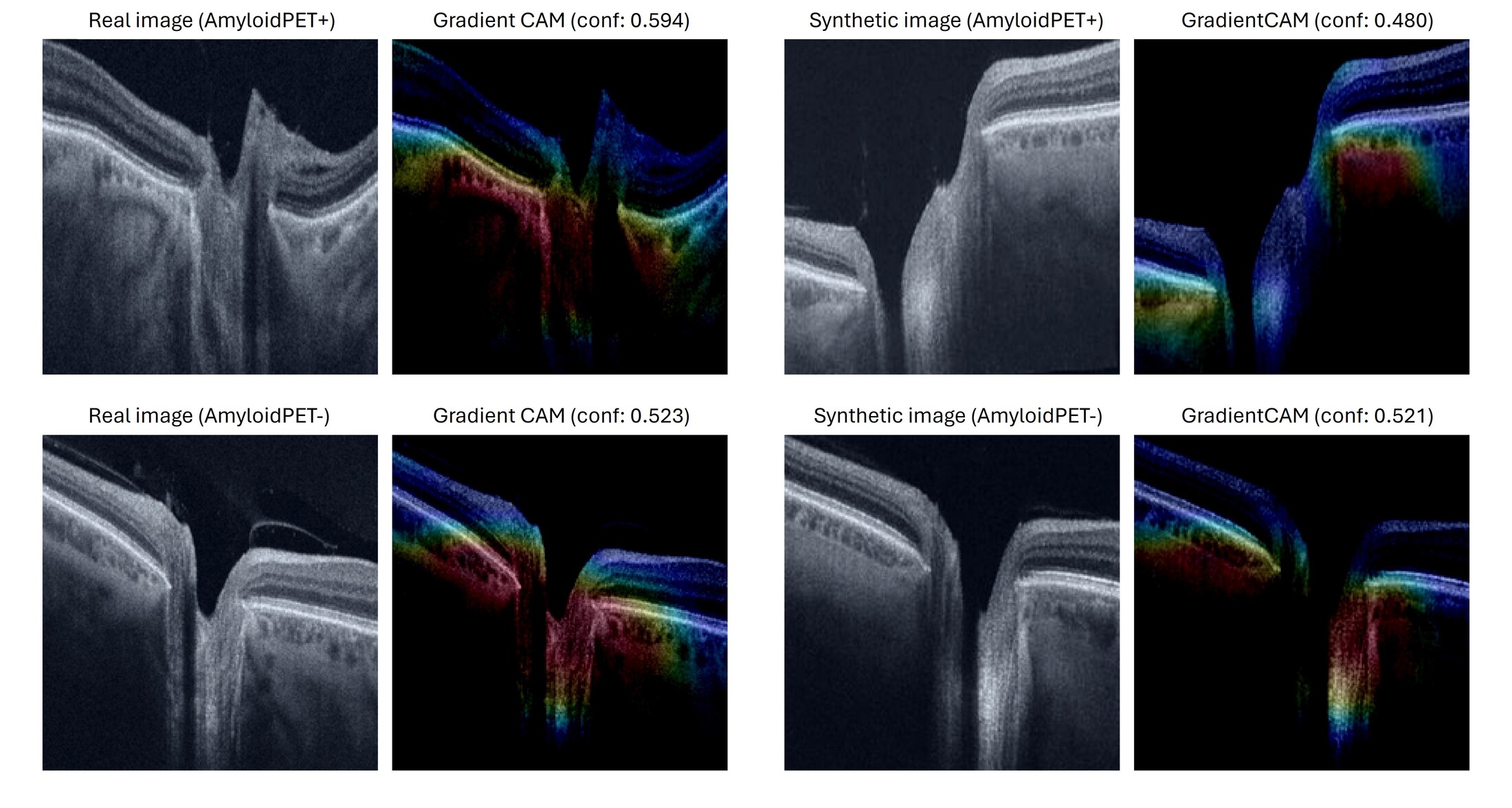}
         \caption{ }
     \end{subfigure}
     \begin{subfigure}[b]{0.48\textwidth}
         \centering
         \includegraphics[width=\textwidth]{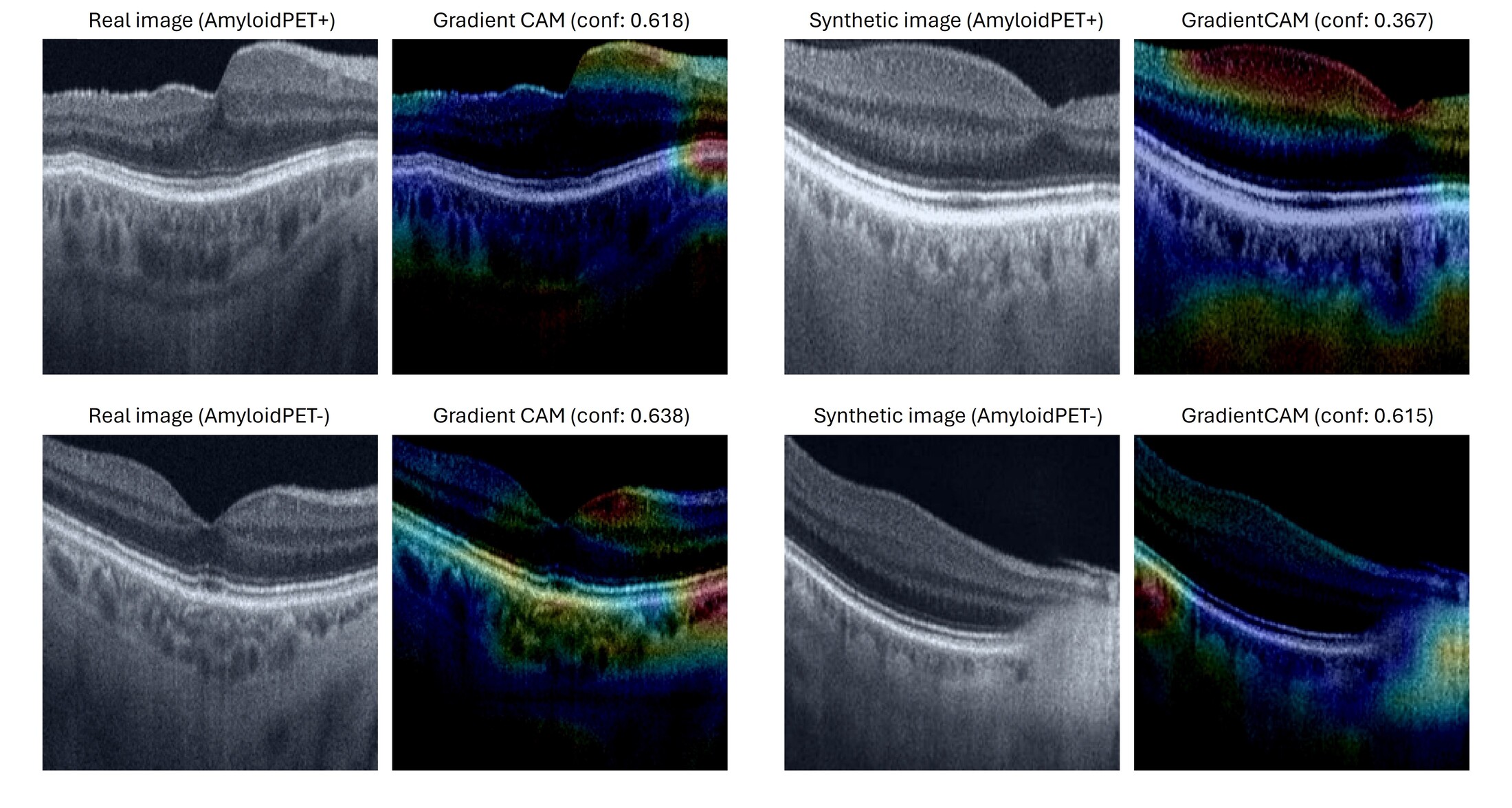}
         \caption{ }
     \end{subfigure}
     \begin{subfigure}[b]{0.48\textwidth}
         \centering
         \includegraphics[width=\textwidth]{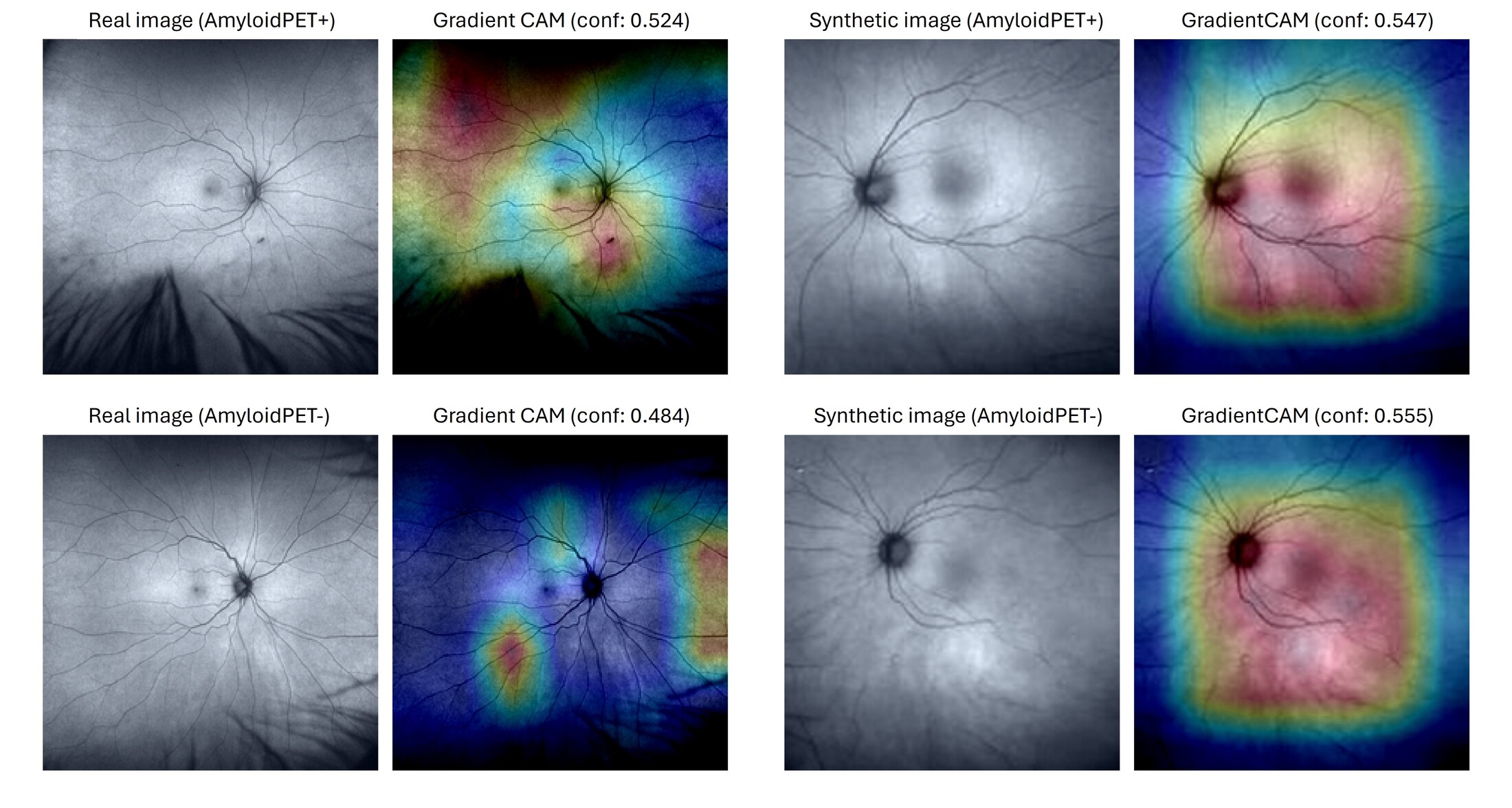}
         \caption{ }
     \end{subfigure}
    \caption{GradCAMs generated for classifiers trained on synthetic images only. The subfigures show pairs of input images and GradCAM heatmaps for OCTA-SMAC (a), OCT-BMONH (b), OCT-BMAC (c), FAF (d). Each subfigure has four images. Left top to bottom: real images for AmyloidPET+ and AmyloidPET-. Right top to bottom: synthetic Images for  AmyloidPET+ and AmyloidPET-. }
    \label{fig:cams_synthetic_classifiers}
\end{figure}